\documentclass[journal,cspaper,compsoc]{IEEEtran}
 
\usepackage{graphicx}
\usepackage{balance}
\usepackage{comment}
\usepackage{cite}
\usepackage{amssymb}
\usepackage[tight,footnotesize]{subfigure}
\usepackage[active]{srcltx}
\usepackage{amsmath}
\usepackage{amsfonts}

\newcommand{\norm}[1]{\left\lVert#1\right\rVert}
\graphicspath{{./figs/}}
\usepackage{color,soul}
\usepackage{eurosym}
\usepackage[plain]{algorithm}
\usepackage{algorithmic}
\usepackage{multicol}
\usepackage{xcolor}
\UseRawInputEncoding
\usepackage{esvect}
\usepackage{bm}
\usepackage{dsfont}

\begin{document}

\title{Distributed Estimation and Control for Jamming an Aerial Target with Multiple Agents}

\author{Savvas~Papaioannou,~Panayiotis~Kolios~and Georgios Ellinas %<-

\IEEEcompsocitemizethanks{\IEEEcompsocthanksitem The authors are with the KIOS Research and Innovation Centre of Excellence (KIOS CoE) and the Department of Electrical and Computer Engineering, University of Cyprus, Nicosia, 1678, Cyprus.\protect\\
E-mail:\texttt{\{papaioannou.savvas, pkolios, gellinas\}@ucy.ac.cy}}% <-this % stops a space
\thanks{}}

\markboth{IEEE Transactions on Mobile Computing, 2022, DOI:10.1109/TMC.2022.3207589 (postprint)}%
{Papaioannou \MakeLowercase{\textit{et al.}}}

\IEEEcompsoctitleabstractindextext{%
\begin{abstract}
This work proposes a distributed estimation and control approach in which a team of aerial agents equipped with radio jamming devices collaborate in order to intercept and concurrently track-and-jam a malicious target, while at the same time minimizing the induced jamming interference amongst the team. Specifically, it is assumed that the malicious target maneuvers in 3D space, avoiding collisions with obstacles and other 3D structures in its way, according to a stochastic dynamical model. Based on this, a track-and-jam control approach is proposed which allows a team of distributed aerial agents to decide their control actions online, over a finite planning horizon, to achieve uninterrupted radio-jamming and tracking of the malicious target, in the presence of jamming interference constraints. The proposed approach is formulated as a distributed model predictive control (MPC) problem and is solved using mixed integer quadratic programming (MIQP). Extensive evaluation of the system's performance validates the applicability of the proposed approach in challenging scenarios with uncertain target dynamics, noisy measurements, and in the presence of obstacles.
\end{abstract}

}

\maketitle
\IEEEdisplaynotcompsoctitleabstractindextext
\IEEEpeerreviewmaketitle

%\color{blue}
\section*{Nomenclature}
\addcontentsline{toc}{section}{Nomenclature}
\begin{IEEEdescription}[\IEEEusemathlabelsep\IEEEsetlabelwidth{$pi_{k|k-1}(x_k|x_{k-1})$}]
	\item[$x_t$] Target state vector at time $t$.
	\item[$y_t $] Target measurement vector at time $t$.
	\item[$u_t $] Target control vector at time $t$.
	\item[$s_t $] Agent state vector at time $t$.
	\item[$v_t $] Agent control vector at time $t$.
	\item[$\omega_t$] Target dynamical model uncertainty.
	\item[$w_t$] Measurement noise.
	\item[$\Phi, \Gamma$] Target/Agent dynamical model parameters.
	\item[$\Delta t$] Sampling time interval.
	\item[$m^\text{agent},m^\text{target}$] Agent and target mass.
	\item[$\eta$] Air resistance coefficient.

	\item[$x_{t^\prime|t}$] Predicted target state at time $t^\prime$ which is computed at time-step $t$.
	\item[$\hat{x}_{t^\prime|t}$] Estimated target state at time $t^\prime$ which is computed at time-step $t$.
	\item[$p_{t|t-1}(x_{t}| y_{1:t-1})$] Predictive distribution of the target state at time $t$.
	\item[$p_{t}(x_{t}| y_{1:t})$] Posterior distribution of the target state at time $t$.
	\item[$f_{t|t-1}(x_{t}|x_{t-1})$] Target transitional density.
	\item[$g_t(y_t|x_t,s_t)$] Measurement likelihood function.
	\item[$\mathcal{S}$] Set of cooperative pursuer agents.
	\item[$R$] Agent sensing range.
	\item[$\mathcal{A}$] Agent sensing profile.
	\item[$\Xi(x_t,s_t)$] Agent detection-and-jamming function.
	\item[$T$] Planning horizon length.
	\item[$X_t$] Target trajectory inside the planning horizon starting at time $t$.
	\item[$U_t$] Target control inputs inside the planning horizon starting at time $t$.
	\item[$\hat{X}_t$] Target estimated trajectory inside the planning horizon starting at time $t$.
	\item[$S_t$] Agent trajectory inside the planning horizon starting at time $t$.
	\item[$V_t$] Agent control inputs inside the planning horizon starting at time $t$.
    \item[$\mathcal{J}_\text{target}$] Target controller objective function.
    \item[$\mathcal{J}_\text{centralized}$] Tracking-and-jamming centralized objective function.
    \item[$\mathcal{J}_\text{agent}$] Tracking-and-jamming distributed objective function.
\end{IEEEdescription}
\!\!\!\!\!\!
\color{black}

\section{Introduction} \label{sec:intro}
%\color{blue}
In recent years, the interest in unmanned aerial vehicle (UAV) technologies for various application areas and industry sectors has soared. UAVs or drones, which were initially utilized as an elite high-tech military technology, have nowadays become widely available and have found widespread use in various domains including emergency response \cite{Papaioannou2021a, Moon2021, Papaioannou2021b, Papaioannou2019, Papaioannou2020} and precision agriculture \cite{Tsouros2019,Radoglou2020} just to name a few. According to the latest Drone Market Report \cite{drone2}, the global market size for drones is predicted to grow to $42.8$B by 2025. The proliferation of consumer drones, however, has also led to the emergence of new threats, with the use of drones for malicious purposes including targeting airports and restricted airspaces \cite{Schneider2019}, attacking critical infrastructures \cite{AC1}, and threatening public safety \cite{Humphreys2015}. These new threats necessitate the research and development of highly efficient, effective, and practical counter-drone systems which can be used to safeguard public safety. In particular, an effective counter-drone system \cite{Wesson2013} must be able to detect, track, and disable the operation of a malicious drone. 
\begin{figure}
	\centering
	\includegraphics[width=\columnwidth]{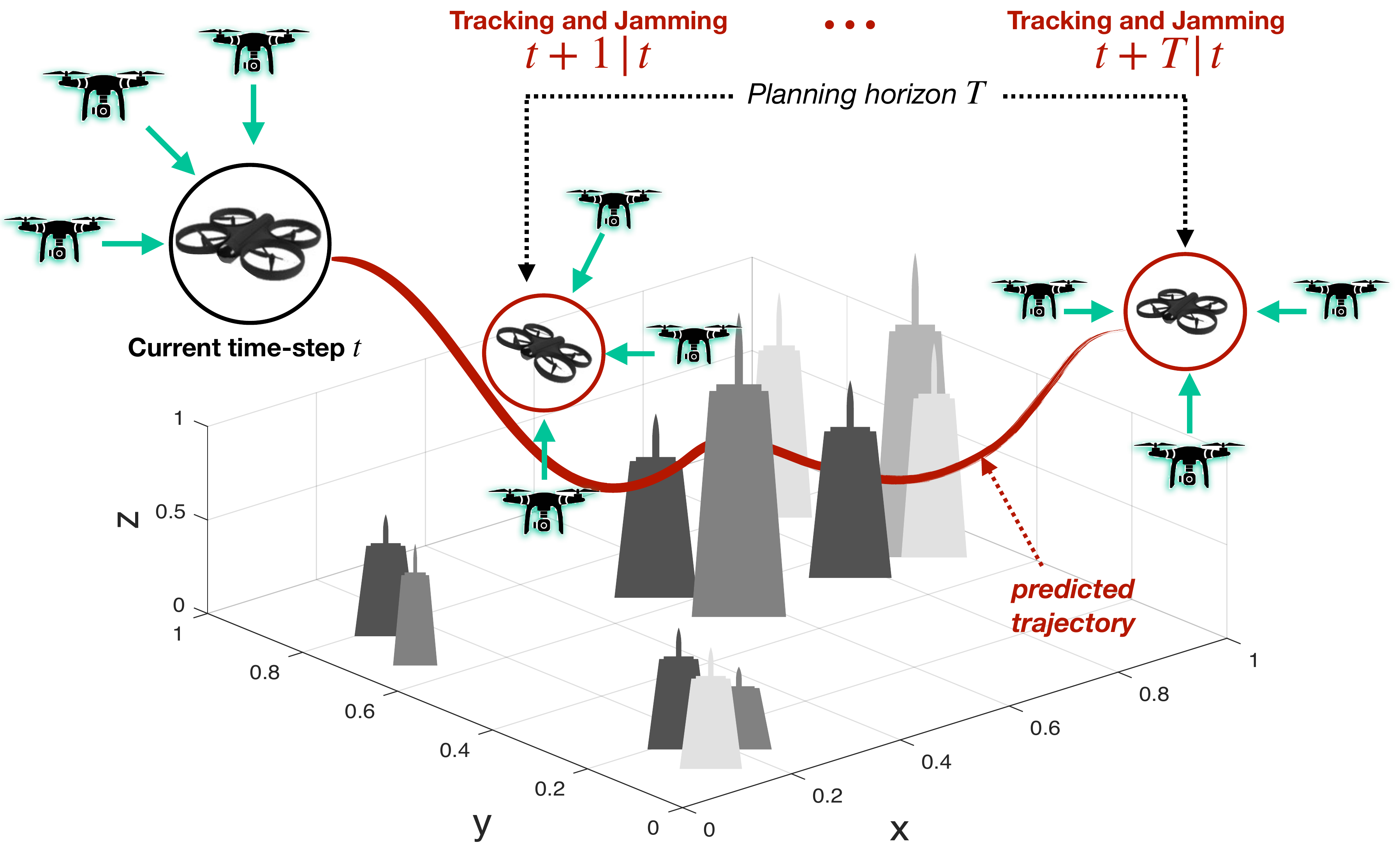}
	\caption{A team of distributed autonomous aerial agents concurrently track and radio-jam a malicious target, while avoiding the jamming interference amongst them and collisions with obstacles in the environment.}	
	\label{fig:sys_arch}
	\vspace{0mm}
\end{figure}

Over the last few years, a variety of counter-drone approaches and systems have been proposed for tackling the problem of malicious drones \cite{Guvenc2018,Loeb2017,Lykou2020}. Anti-drone solutions, however, are still in their infancy and more work is required in order for this technology to reach the required level of maturity. As of today, no adequate solution exists for safeguarding public safety from the threat of malicious drones. Motivated by this necessity, this work proposes a distributed model predictive control (DMPC) \cite{Camponogara2002,Scattolini2009} approach  in which a team of pursuer aerial agents equipped with radio-jamming devices, are collaborating to concurrently track-and-jam a malicious target in 3D space over an extended period of time, with the ultimate purpose of disrupting its navigation circuitry and thus forcing it to execute its fail-safe procedures \cite{Revill2016,Abunada2020} and auto-land. 

Specifically, it is assumed  that the malicious target moves in 3D space according to a stochastic dynamical model and has the ability to maneuver around obstacles and other 3D structures found along its path. The pursuer agents are equipped with a 3D range-finding radar which they use for a) detecting the presence of the malicious target and b) acquiring target measurements. Specifically, the agents use their radar to receive noisy target measurements (i.e., radial distance, azimuth angle, and inclination angle), which they use to recursively estimate the target's state over time. In addition, the pursuer agents are equipped with an omni-directional jamming system \cite{Souli2020} which they use for emitting interference signals at specific wireless bands in an effort to jam the navigation circuitry of the malicious target i.e., disrupting the global navigation satellite system (GNSS) link of the malicious target.

Finally, the agents adapt their decisions and plans on-line over a finite planning horizon in order to better accommodate the collective objective of the team i.e., maximizing the collective tracking-and-jamming performance while at the same time minimizing the jamming interference amongst them. 

\noindent The main contributions of this work are as follows:
\begin{itemize}
\item A distributed estimation and control approach is proposed for the problem of concurrently tracking and radio-jamming a malicious target in 3D space. The problem of tracking-and-jamming is formulated as a constrained model predictive control problem. In this work a team of collaborative pursuer agents decide their control actions online over a rolling finite planning horizon, to achieve uninterrupted radio-jamming and tracking of the malicious target. At the same time the proposed control approach makes sure that the decided control actions over all pursuer agents minimize the induced jamming interference amongst the team. 
\item The problem of concurrent tracking-and-jamming is formulated in noisy and challenging 3D environments i.e., the agents and the malicious target must maneuver around obstacles and other 3D structures to avoid collisions, the malicious target's trajectory is uncertain and thus must be estimated from noisy measurements, and finally the pursuer agents exhibit a 3D range finding sensor with limited sensing range.
\item A mixed integer quadratic mathematical program (MIQP) is derived for the problem tackled, which can be solved exactly using off-the-shelf solvers. The effectiveness of the proposed approach is demonstrated through extensive simulation experiments, addressing scenarios with uncertain target dynamics, noisy measurements, and in the presence of obstacles. 
\end{itemize}

The rest of the paper is organized as follows. Section~\ref{sec:related_work} reviews the state-of-the-art, Section \ref{sec:system_overview} presents the problem definition and system overview, while Section \ref{sec:system_model} introduces the system model. Subsequently, Section \ref{sec:target_control} discusses target maneuvering and target state prediction and estimation, Section \ref{sec:MPC} details the distributed model predictive control approach for tracking-and-jamming, while Section \ref{sec:Evaluation} presents the performance results for a number of experiments. Finally, Section \ref{sec:Conclusion} concludes the work and discusses future research directions.
\begin{figure*}
	\centering
	\includegraphics[width=\textwidth]{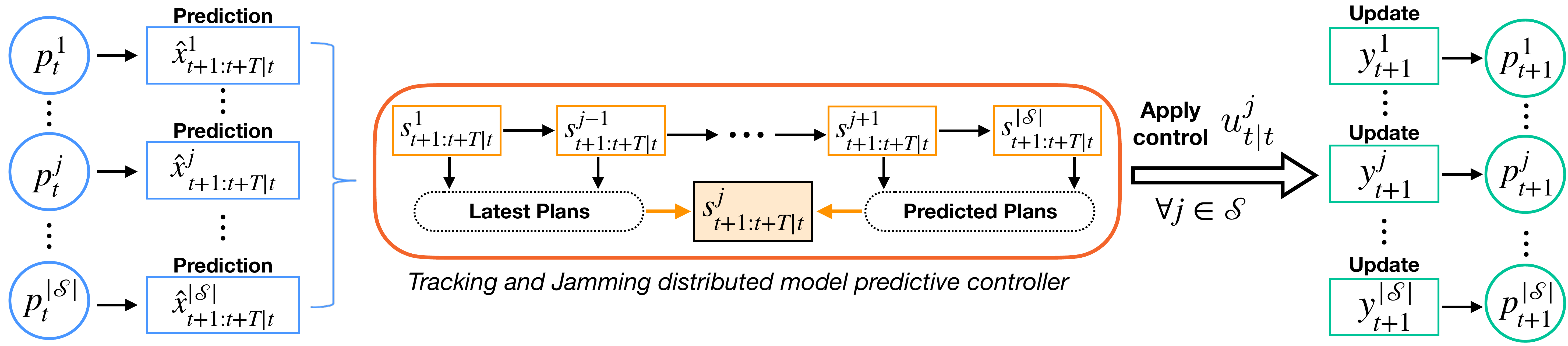}
	\caption{Overview of the proposed tracking-and-jamming system architecture.}	
	\label{fig:arch}
\end{figure*}

\section{Related Work} \label{sec:related_work}
A comprehensive review of the technologies utilized in existing anti-drone systems can be found in \cite{Shi2018}, in which the authors also develop a ground-based anti-drone system capable of detecting, localizing, and radio-jamming a malicious drone. A ground-based radio-frequency (RF) drone detector and jammer system is also proposed in \cite{Abunada2020}. The authors design a drone detection mechanism using the RF control signal exchanged between the drone and its remote controller. Once the malicious drone is detected, a high-power jamming signal is transmitted towards the malicious drone causing it to disconnect from its controller and execute its fail-safe protocols. Furthermore, the  work in \cite{Multerer2017} proposes a low-cost ground jamming system to counteract the operation of small drones. 

Although the ground-based anti-drone systems can be low-cost and effective, they are also associated with the following three major limitations: 
\begin{itemize}
	\item these systems are located at fixed positions on the ground and hence they are only effective when the target resides inside their jamming range. As a consequence, an adversarial target can escape by intelligently navigating outside the system's jamming area. On the other hand in the proposed approach multiple mobile UAV agents with intelligent guidance, navigation, and tracking-and-jamming capabilities can pursue the malicious target, making its escape much harder 
	\item as opposed to mobile anti-drone systems, ground systems appear to be less effective in environments with occlusions and obstacles, which affect their tracking and jamming signals
	\item necessitate the transmission of high-power jamming signals, as opposed to mobile anti-drone systems which can approach the target and use less power for jamming.
\end{itemize} \color{black}

Another recent survey in \cite{Park2021} investigates a wide range of anti-drone technologies, including detection, identification, and neutralization, and proposes guidelines for effective drone defense operations. Various techniques for drone detection, tracking, and interdiction are also discussed in \cite{Guvenc2018} and \cite{Ganti2016}. In \cite{Srigrarom2020} the authors propose a detection and tracking system for small and fast moving drones, whereas in \cite{Dressel2019} the authors use a consumer drone outfitted with antennas and commodity radios to autonomously localize another drone by its telemetry radio emissions. Additionally, the work in \cite{Brust2017} proposes an anti-drone defense system for intercepting and escorting a malicious drone outside a restricted flight zone. In particular, a networked defense drone swarm forms a cluster around the malicious drone to prevent it from entering the restricted airspace. Similarly, the work in \cite{Karras2020} proposes a decentralized formation control anti-drone system, with multiple multi-rotor aerial vehicles cooperating to establish a predefined enclosing formation around the malicious drone. Moreover, many of the existing works (e.g.,~\cite{Brust2017,Karras2020}) assume that a highly accurate drone detection and tracking system is in place, that provides the location of the malicious drone. Instead, in the proposed approach the problem of detecting, tracking, and jamming the malicious drone is investigated jointly. \color{black}

More recently, the works in \cite{Valianti1,Valianti2,Papaioannou1,Papaioannou2}, investigate the problem of target jamming by multiple agents, and they propose centralized and distributed decision and control algorithms which can be used to tackle this problem. More specifically, in \cite{Valianti1} the authors propose a centralized control approach where the jamming power level and mobility control actions of a team of pursuer agents are jointly decided in such way so that the jamming power received at the malicious drone is maximized, whereas in \cite{Valianti2} a distributed control approach is proposed for the same problem. The works in \cite{Valianti1,Valianti2}, however, tackle the problem in obstacle-free 2D environments with simplified discrete kinematic models. In addition, compared to the proposed approach, in which the problem is solved optimally over a rolling finite planning horizon, the approaches presented in \cite{Papaioannou1,Papaioannou2} are myopic greedy one-step ahead planners \color{black}. 

In summary, compared to the existing state-of-the-art techniques, in this work a novel distributed estimation and control approach is proposed for the problem of concurrently tracking-and-jamming a malicious target with a team of distributed mobile agents. The problem is investigated in challenging conditions i.e., 3D environment with obstacles, noisy dynamic and measurement models, and limited sensing range for target detection. Distributed model predictive control (DMPC) is used as the multi-agent control method and distributed stochastic filtering as the target state estimation method. The proposed approach enables the pursuer agents to optimally decide their control actions online over a finite planning horizon, so as to maximize the team's collective objective, while at the same time taking into consideration jamming interference constraints amongst them. Finally, the proposed formulation can be solved exactly and efficiently using off-the-shelf MIQP solvers.

\section{Problem Definition and System Overview}\label{sec:system_overview}
An overview of the proposed system architecture is illustrated in Fig. \ref{fig:arch}. As shown in the figure, each agent $j \in \mathcal{S}$ maintains a probability distribution $p^j_t=p^j_t(x_t|y^j_{1:t})$ which accounts for the uncertainty in the target's state $x_t$ (i.e., the position and velocity) given target measurements $y^j_{1:t}$ up to time $t$. The agents use stochastic filtering to recursively compute in time the posterior distribution of the target state using the time-prediction and measurement-update steps as shown in the figure. More specifically, each agent $j$ uses its current belief of the target state in order to make predictions on the future target trajectory $\hat{x}^j_{t+\tau+1|t}, \tau \in [0,..,T-1]$ over a rolling finite planning horizon of length $T$ time steps, where the notation $t^\prime|t$ indicates the predicted target state at time $t^\prime$ which is made at time $t$. Subsequently, the predicted target trajectories are fused and are used in the distributed model predictive tracking-and-jamming controller as shown in the figure. 

The problem of concurrently tracking-and-jamming a malicious drone with multiple pursuer agents is formulated as a distributed model predictive control (DMPC) problem \cite{Camponogara2002,Scattolini2009}, where at each sampling time $t$ the agent control actions $V^j_t = \{v^j_{t|t},\ldots,v^j_{t+T-1|t}\}, \forall j$ are obtained over a rolling-horizon of length $T$ time steps, by solving a finite horizon open-loop optimal control problem using the current state of the agent $s^j_{t|t}$ as the initial state. The first control action $v^j_{t|t}$ in the sequence is then applied to agent $j$ and the optimization is repeated for the next sampling time. Once agent $j$ applies its first control action in the sequence $v^j_{t|t}$, it moves to its new state where it receives the new target measurement $y^j_{t+1}$ and computes the posterior distribution of the target state $p^j_{t+1}$. The local densities are then fused using covariance intersection and the final estimated target state is extracted from the fused density. The agents then use the fused target density to make predictions for the next planning horizon.

The distributed concurrent tracking-and-jamming controller shown in Fig. \ref{fig:arch} follows a sequential processing procedure \cite{Richards2007}, where each agent $j$ computes its own plan $s^j_{t+\tau+1|t}, \tau \in [0,..,T-1]$ sequentially, one after the other. To compute their plans, the agents collaborate by exchanging information in a sequential coordinated fashion where each agent $j$: (a) receives the latest plans from all agents $i \in [1,..,j-1]$ earlier in the sequence which have already computed their plans and (b) receives the predicted plans from all agents $i \in [j+1,..,|\mathcal{S}|]$ later in the sequence, which have not yet computed their latest plans. %In that sense, the first agent in the sequence will receive the predicted plans of all agents $i \in [2,..,|\mathcal{S}|]$ that have been generated in the previous time step, the second agent in the sequence will then receive the latest computed plan of the first agent and the predicted plans from all agents later in the sequence, and so on.  

Thus, with the proposed tracking-and-jamming controller the agents collaborate in computing their plans, such that the collective tracking-and-jamming objective over the planning horizon is maximized, and at the same time the jamming interference constraints are satisfied (i.e., the jamming interference amongst the team of pursuer agents is minimized).  To summarize, the problem addressed in this work can be stated as follows: 
\textit{Find the optimal control actions $V^j_t =\{v^j_{t+\tau|t} : \forall \tau \in [0,..T-1]\}$ for all agents $j \in \mathcal{S}$, such that the collective task of concurrently tracking-and-jamming the malicious target is maximized, while at the same time the jamming interference between the agents is minimized}.\color{black}

\section{System Model} \label{sec:system_model}
%\color{blue}
A proof-of-concept real-world implementation of a single UAV-based pursuer counter-drone system with detection, tracking and jamming capabilities was demonstrated in our previous work \cite{Souli2020}. This enabled us to gain key-insights in this application-domain, and propose in this work a distributed estimation and control framework for jamming an aerial target with multiple UAV agents. Based on the experiences we have gained from the implementation of a real-world counter-drone system, in this section we outline our modeling assumptions for the problem tackled in this work.
\color{black}
\subsection{Target Dynamical Model}\label{ssec:Target_dynamics}
%\color{blue}
In this work, it is assumed that a single aerial target is maneuvering in three dimensional space according to the following stochastic, state-space \cite{li2003survey,Luis2020} model:
\begin{equation} \label{eq:target_dynamics}
    x_{t} = \Phi x_{t-1} + \Gamma u_{t-1} + \omega_{t-1}
\end{equation}

\noindent where $x_t = [\text{x},\dot{\text{x}}]_t^\top \in \mathbb{R}^6$ denotes the state of the target at time $t$ which consists of position $\text{x}_t=[\text{p}_x, \text{p}_y, \text{p}_z]_t \in \mathbb{R}^3$ and velocity $\dot{\text{x}}_t = [\nu_x,\nu_y,\nu_z]_t \in \mathbb{R}^3$ components in 3D Cartesian coordinates. Furthermore, it is assumed  that the target can change its course and speed via input force $u_t = [\text{u}_x, \text{u}_y, \text{u}_z]_{t}^\top \in \mathbb{R}^3$, and that $\omega_t$ is a Gaussian process noise which represents the model uncertainty \cite{Blackman1999}, and which is distributed according to a zero mean multivariate Gaussian distribution with covariance matrix $Q$ i.e., $\omega_t \sim \mathcal{N}(0,Q)$. Moreover, it is assumed that the target's initial state is distributed according to $\mathcal{N}(\mu_0,P_0)$. The matrices $\Phi$ and $\Gamma$ are further given by:
\begin{equation}\label{eq:PhiGamma}
\Phi = 
\begin{bmatrix}
    \text{I}_{3\times3} & \Delta t \cdot \text{I}_{3\times3}\\
    \text{0}_{3\times3} & \phi \cdot \text{I}_{3\times3}
   \end{bmatrix},
\Gamma = 
\begin{bmatrix}
    \text{0}_{3\times3} \\
     \gamma \cdot \text{I}_{3\times3}
   \end{bmatrix}
\end{equation}

\noindent where $\Delta t$ denotes the sampling time interval, $\text{I}_{3\times3}$ and $\text{0}_{3\times3}$ are the identity matrix and zero matrix of dimension $3 \times 3$, respectively, and parameters $\phi$ and $\gamma$ are given by $\phi =  (1-\eta)$ and $\gamma = \frac{\Delta t}{m^{\text{target}}}$, where $\eta$ is used to model the air resistance coefficient and $m^{\text{target}}$ denotes the mass of the target.
\color{black}

\subsection{Agent Dynamical Model} \label{ssec:Agent_dynamics}
A team $\mathcal{S}=\{s^1,\ldots,s^{|\mathcal{S}|}\}$ of pursuer mobile agents operate inside the 3D surveillance region with the ultimate purpose of tracking-and-jamming the target. The total number of pursuer mobile agents is assumed to be known, fixed, and equal to $|\mathcal{S}|$, where the operator $|.|$ denotes the set's cardinality. The agent dynamics are given by the following linear, discrete-time dynamical model:
\begin{equation} \label{eq:agent_dynamics}
    s^j_{t} = \Phi s^j_{t-1} + \Gamma v^j_{t-1}, ~\forall j \in [1,\ldots,|\mathcal{S}|]
\end{equation}

\noindent where $s_t = [\text{s},\dot{\text{s}}]_t^\top \in \mathbb{R}^6$ denotes the state of the agent at time $t$ which consists of position $\text{s}_t=[\text{p}_x, \text{p}_y, \text{p}_z]_t \in \mathbb{R}^3$ and velocity $\dot{\text{s}}_t = [\nu_x,\nu_y,\nu_z]_t \in \mathbb{R}^3$ components in 3D Cartesian coordinates, and $v_t \in \mathbb{R}^3$ denotes the applied control input at time $t$. The matrices $\Phi$ and $\Gamma$ are given by Eq. \eqref{eq:PhiGamma}. Without loss of generality, it is assumed in this work that all pursuer agents have a mass equal to $m^{\text{agent}}$ and, in general, the pursuer agents have equal capabilities, although this assumption is not strict and can be relaxed. \color{black} With the proposed approach we aim to find the optimal mission-level agent control inputs according to the linear dynamical model of Eq. \eqref{eq:agent_dynamics}, essentially generating the reference trajectory to be executed by the on-board UAV controller \cite{Cowling2007,Fradkov2005}, which can take into account the UAV's low-level controls and aerodynamic behavior.\color{black}

\subsection{Agent Sensing and Jamming Model} \label{ssec:agent_sensing}
%\color{blue}
The pursuer agents are equipped with an onboard omnidirectional active-sensing 3D radar and jamming system \cite{Souli2020} which they use in order to detect the aerial target, acquire target measurements, and transmit power \cite{Abunada2020} to the target, thus radio-jamming its navigation circuitry. In this work, it is assumed that the task of uninterrupted radio jamming will force the target to execute its fail-safe protocols \cite{Revill2016} and thus terminate its mission. This aforementioned fail-safe functionality assumption is motivated by the fact that the majority of non-military grade drones i.e., off-the-shelf consumer drones nowadays implement a certain set of fail-safe procedures (e.g., the auto-land safety protocol when the drone looses its navigation link or its battery is running critically low or the return-to-home fail-safe protocol which is activated when the drone looses its connection with the remote controller) due to increased concerns for public safety. The auto-land fail-safe protocol was indeed tested and verified experimentally through our previous work in~\cite{Souli2020} and is also documented in other related works~\cite{Revill2016, Abunada2020}. We should mention here that the proposed approach can be extended with a directional jamming antenna with slight modifications. \color{black} The agent's active-sensing 3D radar/jamming system exhibits the following characteristics.

\subsubsection{Sensing Profile} Agent $j$ with state $s^j_t$ exhibits a radio sensing profile $\mathcal{A}^j$ in 3D which is modeled as a sphere with Cartesian coordinates $(x_t,y_t,z_t)$ given by:
\begin{align}
    x^j_t &= \text{p}^j_{t,x} + R^j \sin(\theta)\cos(\phi) \notag \\
    y^j_t &= \text{p}^j_{t,y} + R^j \sin(\theta)\sin(\phi) \notag \\
    z^j_t &= \text{p}^j_{t,z} + R^j \cos(\theta) \notag
\end{align}

\noindent where $ \theta \in [0, \pi]$, $\phi \in [0, 2\pi]$, and $\text{s}^j_t =Hs^j_t=[\text{p}^j_x,\text{p}^j_y,\text{p}^j_z]_t^\top$ is the agent's position at time $t$, with $H$ being a matrix which extracts the position coordinates from the agent's state vector $s^j_t$. Finally, $R^j$ is the radius of the sensing profile and denotes the agent's radar and jamming range. Thus, a target with position coordinates $Hx_t$ resides inside agent's $j$ sensing range (i.e., notation $Hx_t \in \mathcal{A}^j$ is used) when $\norm{Hx_t - Hs^j_t}_2 \le R^j$.

\subsubsection{Measurement Acquisition Model} Each pursuer mobile agent $j$ uses its active-sensing radar system to acquire relative 3D target measurements \cite{Taek2001} at every time step, composed of radial distance $\rho$, azimuth angle $\theta$, and inclination angle $\phi$, i.e., $y_t=[\rho,\theta,\phi]^\top_t \in \mathbb{R}^3$, according to the measurement model:
\begin{equation} \label{eq:measurement_model}
    y^j_{t} = h(x_{t},s^j_{t}) + w^j_{t}
\end{equation}

\noindent with $h(x_{t},s^j_{t})$ given by:
\begin{equation} 
    \left[ \norm{H x_{t}-H s^j_{t}}_2, ~ \tan^{-1}(\frac{\Delta_y}{\Delta_x}), ~ \tan^{-1}(\frac{\sqrt{\Delta_x^2+\Delta_y^2}}{\Delta_z}) \right]^\top \notag
\end{equation} 
where $\Delta_x$, $\Delta_y$, and $\Delta_z$ are the differences between the target's and the agent's $x, y$, and $z$ position coordinates respectively, and $w^j_{t} \sim \mathcal{N}(0,\Sigma)$ denotes the zero mean Gaussian measurement noise with covariance matrix $\Sigma$, which models the radar's measurement acquisition process imperfections \cite{Taek2001,Blackman1999}. It should be noted here that target measurements are received only when the target is detected inside the agent's sensing range.
\color{black}

\subsubsection{Target Detection and Jamming Model}

The pursuer agents use their onboard active-sensing 3D radar and jamming system to detect-and-jam the target. More specifically, it is assumed that a target detection can occur only when the target resides inside the agent's radar sensing range $R$. When this happens, the agent acquires target measurements as previously discussed. At the same time, when the target resides inside the agent's sensing range $R$, the target can also be jammed. It should be noted here that the processes of target detection and jamming do not affect each other i.e., they operate on different channels. \color{black}
Hence, it is important to emphasize that the target detection and jamming occurs concurrently when the target resides inside the agent's sensing profile $\mathcal{A}$. More specifically, a target detection-and-jamming event $\Xi^j_t = \Xi^j_t(x_t,s^j_t)$, by agent $j$ with state $s^j_t$ for a target with state $x_t$ occurs when: 
\begin{equation}\label{eq:sensing_model}
 \Xi^j_t(x_t,s^j_t) = 
  \begin{cases} 
   1 & \!\!\!\!\!\!~~~\text{when } Hx_t \in \mathcal{A}^j \\
   0 & \!\!\!\!\!\!~~~ \text{o.w } \notag
  \end{cases}\!\!\!\!\!\!\!\!\!\!\!\!
\end{equation}
\subsection{Obstacle Model} \label{ssec:obstacles}
The obstacles inside the surveillance area are represented in this work as rectangular cuboids of various sizes. A rectangular cuboid is a convex hexahedron in the three dimensional space which exhibits six rectangular faces, where each pair of adjacent faces meets in a right angle. A 3D point $\mathbf{x}=[x,y,z]^\top \in \mathbb{R}^3$ that belongs to the cuboid $\mathcal{C}$ satisfies the following system of linear inequalities:
\begin{align}
    a_{11}x + a_{12}y & + a_{13}z  \le b_1 \notag \\
    a_{21}x + a_{22}y & + a_{23}z  \le b_2 \notag \\
    \vdots & \notag \\
    a_{n1}x + a_{n2}y & + a_{n3}z  \le b_n \notag
\end{align} 
where $n=6$ is the total number of faces which compose the cuboid $\mathcal{C}$, $\boldsymbol{\alpha_i} = [a_{i1},a_{i2},a_{i3}]$ is the outward normal vector of the $i^\text{th}$ face of the cuboid, and $b_i$ is a constant resulting from the dot product between $\boldsymbol{\alpha_i}$ and a known point on the plane that contains the $i^\text{th}$ face. In more compact form, the expression above can be written as $A \mathbf{x} \le B$, where $A$ is a $n \times 3$ matrix, $\mathbf{x}$ is a $3 \times 1$ column vector, and $B$ is a $n \times 1$ column vector. That said, an obstacle in the environment is avoided by an agent $j$ (or the target) with state $s^j_t$ when the following condition is true:
\begin{equation}
    \exists~ i \in [1,..,n]: \text{dot}(\boldsymbol{\alpha_i}, H s^j_t)> b_i
\end{equation}

\noindent where $\text{dot}(\alpha,\beta)$ denotes the dot product between the vectors $\alpha$ and $\beta$. \color{black} It should be mentioned here that the aforementioned obstacle avoidance technique is not limited only to cuboid-like obstacles and can be used for any kind of convex polyhedra. On the other hand, one way to handle a non-convex obstacle, is by embedding it into the minimum bounding convex set. Although this approach would generate conservative collision avoidance trajectories, it fits within the proposed MIQP framework which has the potential to be solved to optimality with existing optimization tools. Finally, as discussed in Sec. \ref{sec:MPC}, since the agent trajectories are computed at each time-step in a receding horizon fashion, dynamic obstacle detection functionality can be easily integrated into the proposed controller. Therefore, information regarding the obstacles do not need to be known a-priori; and can be obtained in an online fashion and processed within each successive planning horizon.
\color{black}

\section{Target Trajectory} \label{sec:target_control}
The target's planned trajectory over a finite planning horizon of length $T$ is denoted as $X_t = \{x_{t+\tau+1|t} : \tau \in [0,..,T-1] \}$, where the notation $x_{t^\prime|t}$ is used to denote the predicted target state at time $t^\prime$ generated when the target is actually at time $t$. The  target trajectory $X_t$ is the solution to the MIQP problem (P1), as discussed next. Essentially, it is assumed in this work that the target's plan is to move as quickly as possible to a designated goal region, while avoiding the obstacles in its path. This is formulated as a rolling-horizon model predictive control problem with linear and binary constraints and is solved using mixed integer quadratic programming (MIQP) by the target's controller.

Problem (P1) finds the target control inputs $U_t =\{u_{t+\tau|t} : \tau \in [0,..,T-1]\}$ which guide the target towards a known designated goal region $\mathcal{G}$, which is represented by a cuboid $\mathcal{C}_\mathcal{G}$ with centroid $\mathcal{G}_o \in \mathbb{R}^3$. To achieve this, the target's controller minimizes the objective function:
\begin{align} \label{eq:mission_objective}
    &\underset{U_t}{\min} ~\mathcal{J}_\text{target}(X_t,U_t) = \|Hx_{t+T|t}-\mathcal{G}_o\|^2_2 \\ &
    ~~~~~~~~~~~~~~~~~~~~~~~+ \lambda \sum_{\tau=1}^{T-1} \|u_{t+\tau|t}-u_{t+\tau-1|t}\|^2_2 \notag
\end{align}

\noindent where the first term drives the predicted target position at time-step $t+T|t$ towards the goal region and the second term in the objective function, i.e., $\sum_{\tau=1}^{T-1} \|u_{t+\tau|t}-u_{t+\tau-1|t}\|^2_2$, is used to minimize the deviations between consecutive control inputs, thus leading to smooth trajectories which can be followed by the drone's controller. The term $\lambda$ is a tuning weight which controls the emphasis given to trajectory smoothing. The value of $\lambda$ can be fitted to data or can be tuned to match the desired trajectory behavior. \color{black}
\begin{algorithm}
\begin{subequations}
\begin{align}
&\hspace*{-3mm}\textbf{Problem (P1)}: \texttt{Target Maneuvering} &  \nonumber\\
& \hspace*{-3mm}~~~~~~~\underset{U_t}{\min} ~\mathcal{J}_\text{target}(X_t,U_t) &\label{eq:objective_P1} \\
&\hspace*{-3mm}\textbf{subject to} ~ \tau \in [0,..,T-1] \textbf{:}  &\nonumber\\
&\hspace*{-3mm} x_{t+\tau+1|t} = \Phi x_{t+\tau|t} + \Gamma u_{t+\tau|t} & \hspace*{1mm} \forall \tau \label{eq:P1_1}\\
&\hspace*{-3mm} A_{\psi,l} H x_{t+\tau+1|t} > B_{\psi,l} - M z_{\tau,\psi,l} &\hspace*{1mm} \forall \tau,\psi, l \label{eq:P1_2}\\
&\hspace*{-3mm} \sum_{l=1}^L z_{\tau,\psi,l} \le L-1 &\hspace*{1mm} \forall \tau,\psi \label{eq:P1_3}\\
&\hspace*{-3mm} z_{\tau,\psi,l} \in \{0,1\} &\hspace*{1mm} \forall \tau,\psi,l \label{eq:P1_4}\\
&\hspace*{-3mm} x_{t|t} = x_{t|t-1} & \hspace*{1mm}\label{eq:P1_5}\\
&\hspace*{-3mm} |\dot{\text{x}}_{t+\tau+1|t}| \le \text{v}^\text{target}_\text{max} &\hspace*{1mm} \forall \tau \label{eq:P1_6}\\
&\hspace*{-3mm} |u_{t+\tau+1|t}| \le u^\text{target}_\text{max} &\hspace*{1mm} \forall \tau \label{eq:P1_7}
\end{align}
\vspace{-0mm}
\end{subequations}
\end{algorithm}

As shown in (P1), Eq. \eqref{eq:P1_1} reflects the target dynamical model and the constraints in Eqs. (\ref{eq:P1_2})-(\ref{eq:P1_3}) define the collision avoidance constraints with all obstacles $\psi \in \Psi$ in the environment. More specifically, the target avoids a collision with an obstacle $\psi \in \Psi$ when:
\begin{equation}
    Hx_{t+\tau+1|t} \notin \mathcal{C}_\psi,~\forall \psi \in \Psi, \forall \tau \in [0,..,T-1]
\end{equation}

\noindent where $\mathcal{C}_\psi$ denotes the cuboid representation of obstacle $\psi$. Suppose that all obstacles to be avoided contain $L=6$ faces, then at time $\tau$ a collision with obstacle $\psi$ is avoided if $\exists ~l \in [1,..,L]: A_{\psi,l} H x_{t+\tau+1|t} > B_{\psi,l}$, where $A_{\psi,l}$ is the outward normal vector to the $l_\text{th}$ face of the cuboid which represents the obstacle $\psi$ and $B_{\psi,l}$ is a constant. In essence $A_{\psi,l}$ and $B_{\psi,l}$ define the equation of the plane which contains the $l_\text{th}$ face of the cuboid (i.e., see Sec. \ref{ssec:obstacles}).  Therefore, the constraint in Eq. \eqref{eq:P1_2} uses $T \times |\Psi| \times L$ binary variables $z_{\tau, \psi, l}$ to check if  the constraint $A_{\psi,l} H x_{t+\tau+1|t} > B_{\psi,l}$ is violated, where $M$ is a big positive constant. Then, constraint \eqref{eq:P1_3} counts the number of times $z_{\tau, \psi, l}$ is activated and ensures that this number is less or equal than $L-1$. Equivalently, the target control inputs are chosen so that the linear system of inequalities $A_{\psi} H x_{t+\tau+1|t} < B_{\psi}, \forall \psi$ is not satisfied, thus achieving obstacle avoidance. Finally, constraints \eqref{eq:P1_6} and \eqref{eq:P1_7} put an upper bound on the target speed and applied force, respectively. In summary, problem (P1) is used by the target to navigate from its position to the goal location while avoiding collisions with the obstacle in its path. %As shown in the sequel, the agent uses the estimated target trajectory computed by (P1) to make its own intelligent decisions for tracking and jamming the target.

\subsection{Target State Prediction} \label{ssec:target_prediction}
In this work, it is assumed that the agents have the capability of predicting the target's trajectory in the near future (i.e., over a finite planning horizon) with some associated uncertainty. Specifically, from an agent's perspective, the target trajectory $X_t = \{x_{t+\tau+1|t}\}$ for $\tau \in [0,..,T-1]$ is a stochastic process, governed by Eq. \eqref{eq:target_dynamics}, where each future state $x_{t+\tau+1|t}$ is distributed according to $x_{t+\tau+1|t} \sim \mathcal{N}(\mu_{t+\tau+1|t},P_{t+\tau+1|t})$ with $\mu_{t+\tau+1|t}$ and $P_{t+\tau+1|t}$ given by:
\begin{subequations}\label{eq:state_prediction}
\begin{align}
    \mu_{t+\tau+1|t} &= \Phi^{\tau+1} \mu_{t} + \sum_{i=0}^{\tau} \Phi^i \Gamma u_{\tau-i} \\
    P_{t+\tau+1|t} &= \sum_{i=0}^{\tau} \Phi^i Q (\Phi^\top)^i + \Phi^{\tau+1} P_{t} (\Phi^\top)^{\tau+1}
\end{align}
\end{subequations}

As discussed next in Sec. \ref{ssec:state_estimation}, when the agents detect the target inside their sensing range, they receive target measurements which they use to compute the posterior distribution of the target state i.e., $\mathcal{N}(\mu_{t},P_{t})$ for the current time-step $t$, using stochastic filtering. We should also note that the agents use Eq. \eqref{eq:state_prediction} to precompute the predicted future distribution of the target's trajectory over the planning horizon i.e., $\mathcal{N}(\mu_{t+\tau+1|t},P_{t+\tau+1|t}), \forall \tau$ and subsequently use the estimated expected target state $\hat{X}_t = \{\hat{x}_{t+\tau+1|t}\}=\{\mu_{t+\tau+1|t}\}, \tau \in [0,..,T-1]$ to plan their future states.

\subsection{Target State Estimation} \label{ssec:state_estimation}
When the pursuer agents detect the target inside their sensing range, they receive target measurements as discussed in Sec. \ref{ssec:agent_sensing}, which they use to improve their estimation of the target state. To achieve this, each mobile agent $j \in \mathcal{S}$ maintains a local Bayes filter \cite{Doucet2000} which is used to recursively estimate and propagate in time the posterior probability distribution $p_{t}(x_{t}|y_{1:t})$ of the target's state $x_{t}$ given all target measurements $y_{1:t}=[y_1,...,y_{t}]$ up to the current time-step $t$. As previously mentioned, the target's state $x_{t}$ follows a first order Markov process, i.e., Eq. \eqref{eq:target_dynamics}, with transition density $f_{t|t-1}(x_{t}|x_{t-1}) = \mathcal{N}(\Phi x_{t-1} + \Gamma u_{t-1}, Q)$, which determines the probability density that the target with state $x_{t-1}$ at time $t-1$ will move to state $x_{t}$ at the next time step. The Bayes filter recursion is given by:
\begin{subequations}\label{eq:bayes_filter}
\begin{align}
    p_{t|t-1}(x_{t}| &y_{1:t-1})= \int f_{t|t-1}(x_{t}|x) p_{t-1}(x|y_{1:t-1}) d_{x} \label{eq:prediction} \\ 
    p_{t}(x_{t}|y_{1:t}) &= \frac{g_t(y_{t}|x_{t},s_{t})p_{t|t-1}(x_{t}|y_{1:t-1})}{\int g_t(y_{t}|x_{t},s_{t})p_{t|t-1}(x_t|y_{1:t-1})d_{x}} \label{eq:update}
\end{align} 
\end{subequations}

\noindent where the function $g_t(y_{t}|x_{t},s_{t})$ is the measurement likelihood function according to Sec. \ref{ssec:agent_sensing} and determines the probability density that the agent with state $s_{t}$ will receive measurement $y_{t}$ from the target with state $x_{t}$, if the target is detected. Therefore $g_t(y_{t}|x_{t},s_{t})$ is given by:
\begin{equation}
    g_t(y_{t}|x_{t},s_{t}) =\left[1-\Xi_t(x_t,s_t)\right]+ \Xi_t(x_t,s_t)\mathcal{N}(h(x_{t},s_{t}),\Sigma).
\end{equation}

\noindent With that said, Eq. \eqref{eq:prediction}, which is referred to as the time-prediction step, uses the transitional density and the prior belief on the target's state to compute the target's predictive density $p_{t|t-1}(x_{t}|y_{1:t-1})$ before receiving the latest measurement $y_t$. Subsequently, Eq. \eqref{eq:update} i.e., the measurement-update step, incorporates the latest target measurement $y_{t}$ and computes the posterior distribution of the target state, from which a point estimate of the target state can be extracted using the expected a posteriori (EAP) or maximum a posteriori (MAP) estimators as discussed in more detail in Sec. \ref{ssec:d_control}. The reader should note that in the preceding discussion the agent index $j$ was omitted for notational clarity.

Each agent $j$ uses the filtering procedure described above to estimate at each time step the posterior distribution of the target state, i.e., $p^j_{t}(x_{t}|y_{1:t}) = \mathcal{N}^j(\mu_{t},P_{t})$, which is used as the initial belief for predicting the target trajectory over the planning horizon as discussed in Sec. \ref{ssec:target_prediction}, where $\mu^j_{t}$ is the EAP point estimate of the target state and $P^j_{t}$ its covariance matrix.

Subsequently, the collection $\{(\mu_{t}^j, P^j_t)\}, j \in \mathcal{S}$ of local target state estimates is combined at each time step using covariance intersection \cite{Julier2017} to produce the fused posterior distribution of the target's state. More specifically, in this step the agents cooperate in a sequential fashion by exchanging their local estimates and computing recursively the fused target state as follows: Agent $1$ first transmits its local estimate $(\mu_{t}^1, P^1_t)$ to agent $2$ which in turn combines the received data with its local data, i.e., $(\mu_{t}^2, P^2_t)$, to compute the fused target distribution $\mathcal{N}(\tilde{\mu}^{1}_t, \tilde{P}^{1}_t)$ between agents $1$ and $2$ as follows:
\begin{subequations} \label{eq:fusion1}
\begin{align}
    (\tilde{P}^{1}_t)^{-1} &= \omega (P^1_t)^{-1} + (1-\omega)(P^2_t)^{-1}\\
    (\tilde{P}^{1}_t)^{-1} \tilde{\mu}^{1}_t &= \omega (P^1_t)^{-1}\mu^1_t + (1-\omega) (P^2_t)^{-1}\mu^2_t
\end{align}
\end{subequations}

\noindent where $\omega \in [0,1]$ is a weighting coefficient \cite{Deng2012} and is found by solving the optimization problem below:
\begin{equation} \label{eq:fusion2}
     \underset{\omega \in [0,1]}{\arg\min} ~ \text{tr} \Big\{\left[ \omega (P^1_t)^{-1} + (1-\omega)(P^2_t)^{-1} \right]^{-1} \Big\}
\end{equation}

\noindent where $\text{tr}\{.\}$ denotes the trace of a matrix. Next, agent $2$ transmits the fused estimate $(\tilde{\mu}^{1}_t, \tilde{P}^{1}_t)$ to agent $3$ and the covariance intersection procedure shown in Eqs. \eqref{eq:fusion1}-\eqref{eq:fusion2} is repeated to produce the fused estimate $(\tilde{\mu}^{2}_t, \tilde{P}^{2}_t)$ by combining agent's $3$ data, i.e., $(\mu^3_t, P^3_t)$. The procedure is repeated until the final fused target distribution $\mathcal{N}(\tilde{\mu}^{|\mathcal{S}|-1}_t, \tilde{P}^{|\mathcal{S}|-1}_t)$ is computed after incorporating the data from the last agent, i.e., $(\mu^{|\mathcal{S}|}_t, P^{|\mathcal{S}|}_t)$. The parameters of the fused posterior target distribution $(\tilde{\mu}^{|\mathcal{S}|-1}_t, \tilde{P}^{|\mathcal{S}|-1}_t)$ are then broadcasted to all agents, which they use as a prior belief for the next time step. We should mention here that in this work we assume that any pre-existing wireless communication technology (e.g., WiFi, ZigBee, Bluetooth) can be used to handle the exchange of information and communication between the agents. Subsequently, low-level communication aspects such as multi-path fading and shadowing effects are assumed to be handled by the wireless communication module.\color{black}

\section{Distributed MPC for Tracking and Jamming} \label{sec:MPC}
This section develops a DMPC algorithm for the cooperative guidance of the team of pursuer UAV agents for the purpose of concurrently tracking-and-jamming the malicious target, while at the same time minimizing the jamming interference among the pursuer agents. Instead of solving a large centralized optimization problem, where all the necessary information is transmitted to a central station which in turn decides the control actions for each agent, this optimization problem is decomposed into smaller sub-problems that can be solved autonomously within each agent. The proposed distributed tracking-and-jamming approach ensures that the control actions taken by one agent are consistent with the control actions of all other agents in the team and that the coupled jamming interference constrains between the pursuer agents are taken into account during the decision making process.

\subsection{Centralized Control}
The centralized version of the problem, as shown in (P2) below, is initially described in a high-level form, followed by a detailed discussion of the proposed DMPC tracking-and-jamming approach.
\begin{algorithm}
\begin{subequations}\!\!\!\!\!\!\!\!
\begin{align}
&\hspace*{-2.5mm}\textbf{Problem (P2)}: \texttt{Centralized MPC} &  \nonumber\\
& \hspace*{-2.5mm}~\underset{V^1_t,..,V^{|\mathcal{S}|}_t}{\max} ~\mathcal{J}_\text{centralized}(\hat{X}_t, S^1_t,..,S^{|\mathcal{S}|}_t,V^1_t,..,V^{|\mathcal{S}|}_t) &\label{eq:objective_P2} \\
&\hspace*{-2.5mm}\textbf{subject to} ~ \forall j \in \{1,..,|\mathcal{S}|\}, \tau \in [0,..,T-1] \textbf{:}  &\nonumber\\
&\hspace*{-2.5mm} s^j_{t+\tau+1|t} = \Phi s^j_{t+\tau|t} + \Gamma v^j_{t+\tau|t} & \hspace*{-13mm} \forall \tau \label{eq:P2_1}\\
&\hspace*{-2.5mm} s^j_{t|t} = s^j_{t|t-1} & \hspace*{-13mm} \label{eq:P2_2}\\
&\hspace*{-2.5mm} s^j_{t+\tau+1|t} \notin \mathcal{C}_\psi & \hspace*{-13mm} \forall \psi \label{eq:P2_22}\\
&\hspace*{-2.5mm} |\dot{\text{s}}^j_{t+\tau+1|t}| \le \text{v}^\text{agent}_\text{max} &\hspace*{-13mm} \forall \tau \label{eq:P2_3}\\
&\hspace*{-2.5mm} |v^j_{t+\tau+1|t}| \le v^\text{agent}_\text{max} &\hspace*{-131mm} \forall \tau \label{eq:P2_4}\\
&\hspace*{-2.5mm} Hs^i_{t+\tau+1|t} \notin \mathcal{A}^j & \hspace*{-13mm} \forall i \ne j \in \mathcal{S} \label{eq:P2_6}
\end{align}
\vspace{-0mm}
\end{subequations}
\end{algorithm}

\subsubsection{Objective Function}
\noindent Problem (P2) optimizes a system-wide objective function i.e., Eq. \eqref{eq:objective_P2}, over the planning horizon of length $T$ with $\tau \in [0,..,T-1]$ for the joint control inputs over all $|\mathcal{S}|$ agents. In (P2), agent's $j$ planned trajectory is denoted as $S^j_t = \{s^j_{t+\tau+1|t} : \forall \tau\}$ and future control inputs as $V^j_t =\{v^j_{t+\tau|t} : \forall \tau\}$. The objective in Eq. \eqref{eq:objective_P2} is also a function of the predicted target trajectory $\hat{X}_t = \{\hat{x}_{t+\tau+1|t} : \forall \tau\}$, since the focus is on finding the control inputs over all pursuer agents that will enable them to track-and-jam the target. The predicted target trajectory $\hat{X}_t$ over the planning horizon has been generated by fusing the local predictions over all agents as discussed in Sec. \ref{ssec:state_estimation}.

In order to track-and-jam the target, the agents must be able to detect the target inside their sensing range as explained in Sec. \ref{ssec:agent_sensing}. This will allow the agents to: (a) emit power to the target in order to jam it and (b) acquire target measurements which they can use to estimate more accurately the target's state, thus improving their tracking performance. Let us assume that the system-wide objective function shown in Eq. \eqref{eq:objective_P2} returns the total number of detection-and-jamming events over the planning horizon for all agents. Thus, whenever the condition $H \hat{x}_{t+\tau+1|t} \in \mathcal{A}^j_{t+\tau+1|t}, \forall \tau,  \forall j$ is true, the value of the objective function is increased. Therefore, in order to maximize the number of detection-and-jamming events over the planning horizon and for all agents it suffices to maximize Eq. \eqref{eq:objective_P2}. From the discussion above, it is clear that the overall system has decoupled agent dynamics and that the objective function in Eq. \eqref{eq:objective_P2} is separable and can be written as:
\begin{equation}
   \sum_{j=1}^{|\mathcal{S}|} \mathcal{J}^j_\text{agent}(\hat{X}_t, S^j_t, V^j_t) \!\!=\!\! \sum_{j=1}^{|\mathcal{S}|}  \sum_{\tau=0}^{T-1}\Xi^j_t(\hat{x}_{t+\tau+1|t},s^j_{t+\tau+1|t})
\end{equation}

\noindent where function $\Xi^j_t(.)$ was defined in Sec. \ref{ssec:agent_sensing}. With this decomposition, each agent decides only its own control inputs when optimizing its local objective. Thus, the agents can work in a distributed fashion optimizing their local objective, while at the same time the system-wide objective is also optimized. Then, Eqs. \eqref{eq:P2_1}-\eqref{eq:P2_2} are constraints imposed by the agent dynamical model assuming a known initial state, Eq. \eqref{eq:P2_22} defines obstacle avoidance constraints, and constraints \eqref{eq:P2_3} and \eqref{eq:P2_4} keep the agent within the allowable speed limit and control input values, respectively.

\subsubsection{Jamming Interference Constraints}
Constraint \eqref{eq:P2_6} ensures that there is no jamming interference between the team of pursuer agents. Agent $j$ is jamming some other agent $i$ at time $t$ when $Hs^i_{t} \in \mathcal{A}_t^j$, i.e., when agent $i$ resides inside the sensing range of agent $j$ at time $t$. Thus, to avoid the interference between the agents it is required that $Hs^i_{t+\tau+1|t} \notin \mathcal{A}^j, \forall i \ne j, \forall j$ or equivalently:
\begin{equation} \label{eq:concave}
     \norm{Hs^i_{t+\tau+1|t} - Hs^j_{t+\tau+1|t}}_2 > R^j,~  \forall \tau, \forall i \ne j, \forall j.
\end{equation}

\begin{figure}
	\centering
	\includegraphics[width=\columnwidth]{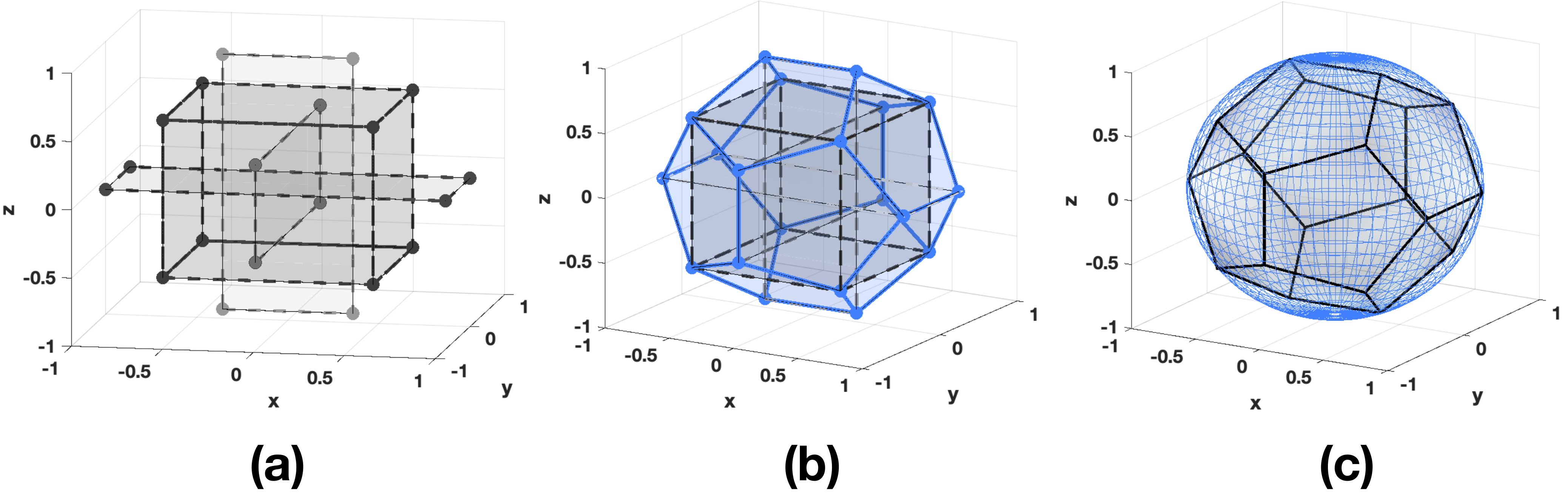}
	\caption{(a)-(b) The $20$ vertices of a regular dodecahedron (centered at the origin with $R=1$) formed from the vertices of a cube and $3$ planes according to Eq. \eqref{eq:dodecahedron_construction}, (c) The agent's sensing area i.e., a dodecahedron with circumradius $R$.}	\label{fig:dodecahedron}
	\vspace{0mm}
\end{figure}

\noindent The constraints in Eq. \eqref{eq:concave}, however, are very challenging to be applied directly, since they create an overall non-convex feasibility region which renders their use impractical, even for specialized solvers. To overcome this problem, the jamming interference constraints of Eq. \eqref{eq:concave} are convexified and approximated by a set of linear constraints which can be handled efficiently by off-the-shelf MQIP solvers. More specifically, the sensing profile of agent $j$, i.e., a sphere with radius $R^j$, is approximated in this work by a regular dodecahedron $\mathcal{D}^j$ with circumradius $R^j$. That is, the circumscribed sphere with radius $R^j$ of the dodecahedron is a sphere that contains the dodecahedron and touches each of the dodecahedron's vertices. A regular dodecahedron centered at the origin exhibits $12$ regular pentagonal faces and $20$ vertices. These $20$ vertices are formed as the union of the vertices of a cube and the vertices of $3$ rectangles on the $yz$, $xz$, and $xy$ planes, with Cartesian coordinates given by (shown in Fig. \ref{fig:dodecahedron}):
\begin{subequations}\label{eq:dodecahedron_construction}
\begin{align}
  \text{Cube}: & ~ \left(\pm \frac{R}{\sqrt{3}},~ \pm \frac{R}{\sqrt{3}},~ \pm \frac{R}{\sqrt{3}}\right)\\
\hspace{-0.5cm}   \text{Rectangle on $yz$}: & ~ \left(0,~ \pm \frac{\varphi R}{\sqrt{3}},~ \pm \frac{R}{\varphi\sqrt{3}}\right)\\
\hspace{-0.5cm}    \text{Rectangle on $xz$}: & ~ \left(\pm \frac{R}{\varphi \sqrt{3}},~ 0,~  \pm \frac{\varphi R}{\sqrt{3}}\right)\\
\hspace{-0.5cm}    \text{Rectangle on $xy$}: & ~ \left(\pm \frac{\varphi R}{\sqrt{3}},~ \pm \frac{R}{\varphi \sqrt{3}},~ 0\right)
\end{align}
\end{subequations}

\noindent where constant $\varphi=\frac{1+\sqrt{5}}{2}$ is the golden ratio. Similarly to the discussion of Sec. \ref{ssec:obstacles}, a point $\mathbf{x}=[x,y,z]^\top \in \mathbb{R}^3$ resides inside the dodecahedron if it satisfies the linear system of $12$ inequalities $D \mathbf{x} \le  C$, where $D$ is a $12 \times 3$ matrix (each row $D_l, l \in [1,..12]$ is the normal to the $l^\text{th}$ face of the dodecahedron) and $C$ is a $12 \times 1$ column vector. With the dodecahedron approximation the interference constraint in Eq. \eqref{eq:concave} can be approximated by a system of $12$ linear inequalities as:
\begin{align}
    Hs^i_{t+\tau+1|t} \notin \mathcal{A}^j & \implies \notag \\
    \exists ~l \in [1,..,12] : & ~D_l (Hs^i_{t+\tau+1|t} - Hs^j_{t+\tau+1|t}) > C_l. \notag
\end{align}

\subsection{Distributed Control} \label{ssec:d_control}
Subsequently, we transform the centralized problem shown in (P2) to the  distributed tracking-and-jamming controller shown in (P3), which can be implemented as an MIQP and solved using off-the-shelf solvers. The problem shown in (P3) is executed at each time step locally within each agent, in a sequential fashion \cite{Richards2007}, where agent $j$ receives the latest plans $s^i_{t+\tau+1|t}$ from all agents $i<j$ earlier in the sequence and also receives the predicted plans $\tilde{s}^i_{t+\tau+1|t} = s^i_{t+\tau|t-1} $ from all agents  $i>j$ later in the sequence, which have not yet computed their latest plans. That said, each agent $j$ optimizes the objective function below:
%\begin{align} \label{eq:mi}
%    &\underset{V_t^j}{\min} ~\mathcal{J}^j_\text{agent}(\hat{X}_t,S^1_t,..,S^{|\mathcal{S}|},V^j_t) = - w_1 \sum_{\tau=0}^{T-1} \tilde{b}^j_{\tau} ~+ \\
%    & w_2 \sum_{\tau=0}^{T-1} \sum_{i \ne j=1}^{|\mathcal{S}|} \tilde{\kappa}^j_{\tau,i} + w_3 \sum_{\tau=1}^{T-1} \|v^j_{t+\tau|t}-v^j_{t+\tau-1|t}\|^2_2 \notag
%\end{align}
\begin{align} \label{eq:mi}
    &\underset{V_t^j}{\min} ~\mathcal{J}^j_\text{agent}(\hat{X}_t,S^1_t,..,S^{|\mathcal{S}|},V^j_t) = - w_1 \sum_{\tau=0}^{T-2} \tilde{b}^j_{\tau} ~+ \\
    & w_2 \sum_{\tau=0}^{T-2} \sum_{i \ne j=1}^{|\mathcal{S}|} \tilde{\kappa}^j_{\tau,i} + w_3 \|s^j_{t+T|t}-\mathcal{G}_o\|^2_2 \notag
\end{align}

\noindent where $\hat{X}_t = \{\hat{x}_{t+\tau+1|t} : \forall \tau\}$ is the expected target trajectory over the planning horizon $t+1|t,..,t+T|t$, $S^{i \ne j}_t$ are the received plans from all other agents, which contain the latest or the predicted plans depending on the position of agent $j$ in the sequence, and $V^j_t =\{v^j_{t+\tau|t} : \forall \tau\}$ are the future control inputs of agent $j$ which are to be optimized. The tracking-and-jamming objective function shown in Eq. \eqref{eq:mi} consists of three terms. The first term is responsible for the detection and jamming events and thus maximizes (i.e., minimizes the negative) the binary variable $\tilde{b}^j_{\tau}$ over the planning horizon (i.e., the binary variable which indicates whether the target with state $\hat{x}_{t+\tau+1|t}$ resides inside agent's $j$ sensing range). In other words, this terms implements the function $\Xi^j_t$ as discussed in Sec. \ref{ssec:agent_sensing}. Subsequently the second term minimizes the jamming interference events inside the planning horizon, by making use of the binary variable $\tilde{\kappa}^j_{\tau,i}$ which indicates if agent $i$ is being jammed by agent $j$ at time-step $t+\tau+1|t$. Finally, the last term is used to drive the agent to the goal region. Parameters $w_1, w_2$, and $w_3$ are tuning weights which determine the preference given to each term.
%Finally, the last term is used to obtain a trajectory with smooth curvature. 
\begin{algorithm}
	\begin{subequations}
		\begin{align}
		&\hspace*{-3.0mm}\textbf{Problem (P3)}: \texttt{Distributed Control} &  \nonumber\\
		& \hspace*{-3.0mm}~~~~~~~\underset{V^j_t}{\min} ~\mathcal{J}^j_\text{agent}(\hat{X}_t,S^1_t,..,S^{|\mathcal{S}|},V^j_t) &\label{eq:objective_P3} \\
		&\hspace*{-3.0mm}\textbf{subject to} ~ \tau \in [0,..,T-1], i\ne j\in [1..|\mathcal{S}|]  \textbf{:}  &\hspace*{-6.5mm}  \nonumber\\
		&\hspace*{-3.0mm} s^j_{t+\tau+1|t} = \Phi s^j_{t+\tau|t} + \Gamma v^j_{t+\tau|t} & \hspace*{-6.5mm} \forall \tau \label{eq:P3_1}\\
		&\hspace*{-3.0mm} D_{l} (H\hat{x}_{t+\tau+1|t}\!\!-\!\!Hs^j_{t+\tau+1|t})+ (M\!\!-\!\!C_l) b^j_{\tau,l}\!\! <\!\! M & \hspace*{-6.5mm} \forall \tau, l \label{eq:P3_2}\\
		&\hspace*{-3.0mm} 12 \tilde{b}^j_{\tau} -\sum_{l=1}^{12} b^j_{\tau,l} \le 0 & \hspace*{-6.5mm} \forall \tau \label{eq:P3_3}\\
		&\hspace*{-3.0mm} D_{l} (Hs^i_{t+\tau+1|t}\!\!-\!\!Hs^j_{t+\tau+1|t})> C_l \!\!-\!\! M \kappa^j_{\tau,i,l} & \hspace*{-6.5mm} \forall \tau, l, i<j \label{eq:P3_4}\\
		&\hspace*{-3.0mm} D_{l} (H\tilde{s}^i_{t+\tau+1|t}\!\!-\!\!Hs^j_{t+\tau+1|t})> C_l \!\!-\!\! M \kappa^j_{\tau,i,l} & \hspace*{-6.5mm} \forall \tau, l, i>j \label{eq:P3_41}\\
		&\hspace*{-3.0mm} \sum_{l=1}^{12} \kappa^j_{\tau,i,l} -11 \le \tilde{\kappa}^j_{\tau,i} & \hspace*{-20.5mm} \forall \tau, i \ne j  \label{eq:P3_5}\\
		&\hspace*{-3.0mm} A_{\psi,l} H s^j_{t+\tau+1|t} > B_{\psi,l} - M z_{\tau,\psi,l} &\hspace*{-6.5mm} \forall \tau,\psi, l \label{eq:P3_6}\\
		&\hspace*{-3.0mm} \sum_{l=1}^L z_{\tau,\psi,l} \le L-1 &\hspace*{-6.5mm} \forall \tau,\psi \label{eq:P3_7}\\
		&\hspace*{-3.0mm} b^j_{\tau,l}, \tilde{b}^j_{\tau}, \kappa^j_{\tau,i,l}, \tilde{\kappa}^j_{\tau,i}, z_{\tau,\psi,l} \in \{0,1\} &\hspace*{-1.2mm} \forall \tau,\psi,l \label{eq:P3_8}\\
		&\hspace*{-3.0mm} s^j_{t|t} = s^j_{t|t-1} & \hspace*{-6.5mm}\label{eq:P3_9}\\
		&\hspace*{-3.0mm} |\dot{\text{s}}_{t+\tau+1|t}| \le \text{v}^\text{agent}_\text{max} &\hspace*{-6.5mm} \forall \tau \label{eq:P3_10}\\
		&\hspace*{-3.0mm} |v^j_{t+\tau+1|t}| \le v_\text{max} &\hspace*{-6.5mm} \forall \tau \label{eq:P3_11}
		\end{align}
		\vspace{-0mm}
	\end{subequations}
\end{algorithm}

In formulation (P3) of the distributed tracking-and-jamming controller, constraint \eqref{eq:P3_1} is imposed by the agent dynamical model discussed in Sec. \ref{ssec:Agent_dynamics}, where constraints \eqref{eq:P3_2}-\eqref{eq:P3_3} are responsible for the target detection and jamming events. More specifically, the inequality $D_{l} (H\hat{x}_{t+\tau+1|t}-Hs^j_{t+\tau+1|t}) < C_l$ becomes true when the target's location resides inside the negative half-space created by the plane containing the $l^\text{th}$ face of the dodecahedron that approximates the agent's spherical sensing range. When this happens, the corresponding binary variable $b^j_{\tau,l}$ becomes $1$ to satisfy the constraint. Otherwise, $b^j_{\tau,l}=0$ and constraint \eqref{eq:P3_2} is valid with the addition of a large positive constant $M$ (i.e., big-$M$). Subsequently, the binary variable $\tilde{b}^j_{\tau}$ in Eq. \eqref{eq:P3_3} is activated when the system of linear inequalities is satisfied for all $12$ faces of the dodecahedron, i.e., $D_{l} (H\hat{x}_{t+\tau+1|t}-Hs^j_{t+\tau+1|t}) < C_l, \forall l$ which indicates that the target's location $Hx_{t+\tau+1|t}$ resides inside agent's $j$ sensing range at time $t+\tau+1|t$. Next, Eqs. \eqref{eq:P3_4}-\eqref{eq:P3_5} define the radio-jamming interference constraints amongst the team of agents. In this case, the binary variable $\kappa^j_{\tau,i,l}$ is activated when agent $i$ resides inside the negative half-space created by the plane containing the $l^\text{th}$ face of the dodecahedron which approximates agent's $j$ sensing range. When $\sum_{l=1}^{12} \kappa^j_{\tau,i,l}$ equals $12$, then agent $i$ is being jammed by agent $j$ at time-step $t+\tau+1|t$ (i.e., agent $i$ is inside the dodecahedron), a scenario that must be avoided. For this reason, the binary variable $\tilde{\kappa}^j_{\tau,i}$ shown in Eq. \eqref{eq:P3_5} is minimized, so that $\sum_{l=1}^{12} \kappa^j_{\tau,i,l} < 12$ and the jamming interference is avoided i.e., agent $i$ is outside agent's $j$ sensing range. Finally, constraints \eqref{eq:P3_6}-\eqref{eq:P3_7} ensure obstacle avoidance similarly to (P1) in Sec. \ref{sec:target_control}, and the rest of the constraints define the decision variables and the limits on the speed and input controls of the agent.

It should also be noted that due to the non-linear measurement model, i.e., Eq. \eqref{eq:measurement_model}, the Bayes filter recursion described in Eq. \eqref{eq:bayes_filter} is implemented in this work as a particle filter \cite{Arulampalam2002,Gustafsson2010} and thus the posterior density at time $t-1$ denoted by $p_{t-1}(x_{t-1}|y_{1:t-1})$ is represented by a set of weighted particles $\{w^{(i)}_{t-1}, x^{(i)}_{t-1}\},~ i \in [1,...,N]$:
\begin{equation}
    p_{t-1}(x_{t-1}|y_{1:t-1}) \approx \sum_{i=1}^N w^{(i)}_{t-1} \delta(x_t-x^{(i)}_{t-1})
\end{equation}

\noindent where $\delta(x)$ denotes the Dirac impulse function and $\sum_{i=1}^N w^{(i)}_{t-1} = 1$. Given the proposal density $q(.|x^{(i)}_{t-1},y_{1:t})$ the posterior distribution of the subsequent time-step $t$ is then approximated by a new set of particles $\{w^{(i)}_{t}, x^{(i)}_{t}\}$ as:
\begin{equation}
    p_{t}(x_{t}|y_{1:t}) \approx \sum_{i=1}^N w^{(i)}_{t} \delta(x_{t}-x^{(i)}_{t})
\end{equation}

\noindent where 
\begin{subequations} \label{eq:particle_filter}
\begin{align}
    x^{(i)}_{t} &\sim q(x^{(i)}_{t}|x^{(i)}_{t-1},y_{1:t}) \\ 
    w^{(i)}_{t} &= \tilde{w}^{(i)}_{t} \left(\sum_i \tilde{w}^{(i)}_{t}\right)^{-1} \\
    \tilde{w}^{(i)}_{t} &= w^{(i)}_{t-1} \frac{g_t(y_{t}|x^{(i)}_{t},s_{t}) f_{t|t-1}(x^{(i)}_{t}|x^{(i)}_{t-1})}{q(x^{(i)}_{t}|x^{(i)}_{t-1},y_{1:t})}.
\end{align}
\end{subequations}

\noindent In this work, for the proposal distribution the transition density is used, i.e., $q(x^{(i)}_{t}|x^{(i)}_{t-1},y_{1:t}) = f_{t|t-1}(x^{(i)}_{t}|x^{(i)}_{t-1})$. Subsequently, each agent computes the expected posteriori (EAP) target state point estimate and its associated covariance matrix as:
\begin{subequations} \label{eq:EAP}
\begin{align}
    \mu_{t} &= \sum_{i=1}^N w^{(i)}_{t} x^{(i)}_{t} \\ 
    P_{t} &= \sum_{i=1}^N w^{(i)}_{t} (x^{(i)}_{t}-\mu_{t})(x^{(i)}_{t}-\mu_{t})^\top.
\end{align}
\end{subequations}

To conclude, each agent implements the Bayes recursion in Eq. \eqref{eq:bayes_filter} as a particle filter, which is used to estimate the posterior distribution of the target state at each time step as discussed above. The agents cooperate by exchanging their local estimates to produce the final target distribution using covariance intersection fusion as discussed in Sec. \ref{sec:target_control}, and use the distributed tracking-and-jamming controller discussed in Sec. \ref{sec:MPC} to plan their trajectories such that the target is optimally tracked-and-jammed over the rolling planning horizon with minimum jamming interference.

\section{Performance Evaluation}\label{sec:Evaluation}
\subsection{Simulation Setup} \label{ssec:sim_setup}

The simulation setup used to evaluate the performance of the proposed approach is based on field tests conducted with the DJI M200 series drone \cite{Souli2020}. In particular we assume that the agents and the target maneuver inside a bounded 3D surveillance area of size $300\text{m} \times 300\text{m} \times 100\text{m}$. The dynamics of the target are given by Eq. \eqref{eq:target_dynamics} with $Q=\text{diag}([1, 1, 1, 0.5, 0.5, 0.5])$, $\eta = 0.2$, $m^\text{target}=3.25$kg, and $\Delta t=1$s. The maximum target speed and control input are $\text{v}^\text{target}_\text{max} = 15$m/s and $u^\text{target}_\text{max}=10$N, respectively. Finally, parameter $\lambda$ in Eq. \eqref{eq:mission_objective} is set at $\lambda = 0.01$. Regarding the agents, their dynamics are given by Eq. \eqref{eq:agent_dynamics}, with $\eta = 0.2$ and $m^\text{agent} = 3.25$kg, for all agents. The covariance $\Sigma$ of the measurement noise is set at  $\Sigma=\text{diag}([0.5, \pi/50, \pi/50])$ and the maximum agent speed and control input are $\text{v}^\text{agent}_\text{max} = 20$m/s and $u^\text{agent}_\text{max}=25$N, respectively. The agent radius is $R=20$m for all agents and parameters $w_1, w_2$, and $w_3$ of the objective function in Eq. \eqref{eq:mi} are $w_1=1, w_2=1$, and $w_3=1$. The obstacles in the environment are modeled as rectangular cuboids according to Sec. \ref{ssec:obstacles}, and the target state estimation was implemented as a particle filter \cite{Arulampalam2002,Gustafsson2010}, with $1500$ particles. Simulations were conducted on a $2$GHz dual core CPU computer running Matlab and the Gurobi solver. \color{black} 

\subsection{Results}
\begin{figure*}
	\centering
	\includegraphics[width=\textwidth]{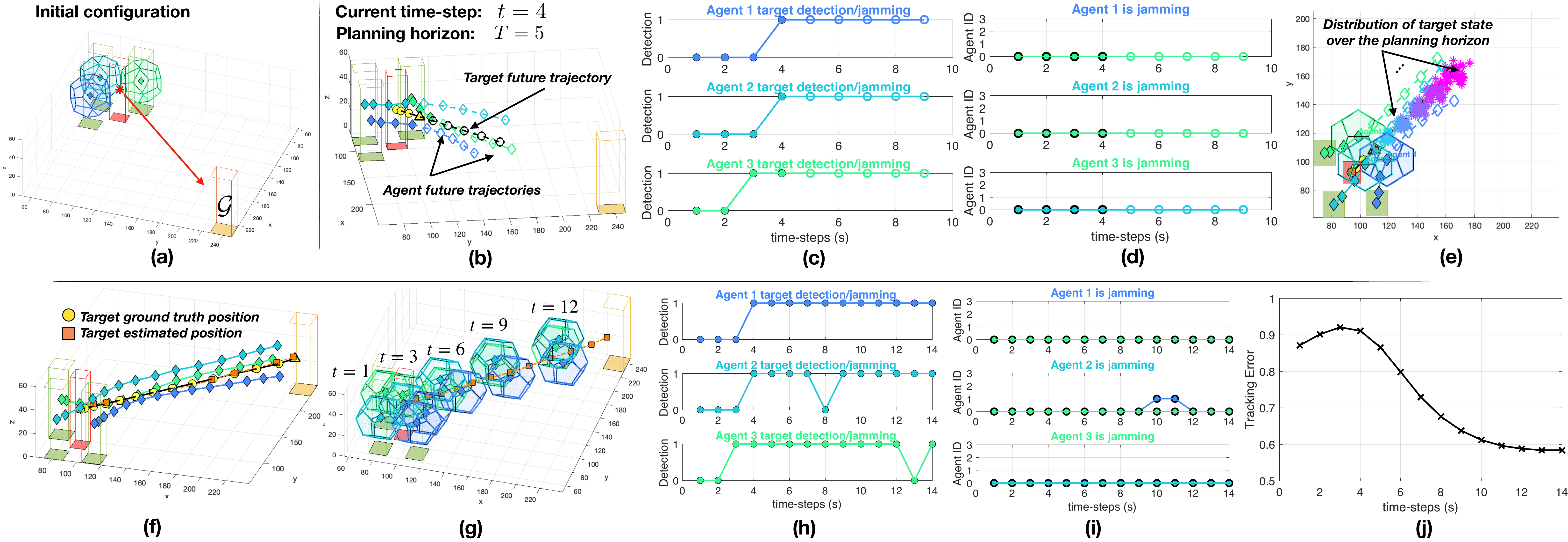}
	\caption{Concurrent tracking-and-jamming scenario with $3$ agents and $1$ target (agents are marked with $\blacklozenge$, with agent $1$, $2$, and $3$ colored blue, turquoise, and green respectively. The ground truth target trajectory (marked with $\bullet$) is colored yellow and the estimated trajectory (marked with $\blacksquare$) is colored orange. The clear shapes  $\diamond$ and $\circ$ indicate the agents and target predicted states, respectively).}	
	\label{fig:fig4}
\end{figure*}

The first experiment aims to demonstrate the proposed approach with a tracking-and-jamming scenario taking place in an obstacle-free environment with $3$ agents and $1$ target. More specifically, in this setup, three agents colored blue, turquoise, and green are spawned at locations $[110,70,30]$, $[80,70,30]$, and $[80,100,30]$, respectively, as shown in Fig. \ref{fig:fig4}(a). The target is initially distributed according to $\mathcal{N}(\mu_0,P_0)$, where $\mu = [95,95,30,3,3,0]$ (shown with a red star) and $P_0=\text{diag}([2, 2, 2, 0.8, 0.8, 0.8])$. The target's guidance controller executes the program shown in (P1), which is responsible for guiding the target to the goal region $\mathcal{G}$ as shown in the figure. Figures \ref{fig:fig4}(b)-(e) show the overall system state at time-step $t=4$ with a planning horizon $T=5$ time steps. More specifically, Fig.\ref{fig:fig4}(b) shows the trajectories of the agents and the target where clear diamonds are used to denote the agent predicted states, and clear circles to denote the target's predicted states over the planning horizon. The executed trajectories are marked with solid shapes. Initially, the target does not reside inside any of the agents' sensing ranges, thus for the initial time steps the target cannot be jammed by any agent as shown in Fig. \ref{fig:fig4}(c). During this time, the agents do not receive any target measurements, since the target is not detected inside their sensing ranges.

Specifically, Fig. \ref{fig:fig4}(c) shows the target detection and jamming events per agent up to the current time-step $t=4$ and over the planning horizon $t+\tau+1|t, \tau \in [0,T-1]$. Similarly, Fig. \ref{fig:fig4}(d) shows the jamming interference events amongst the agents, and Fig. \ref{fig:fig4}(e) shows the system configuration at the current time step in a top-down view, where the distribution of the target state over the planning horizon (as shown) is approximated with a set of particles. The final target and agent trajectories of this experiment are shown in Fig. \ref{fig:fig4}(f). As can be observed, the agents move in such a formation that allows them to surround the target, while at the same time avoiding the jamming interference amongst them. In this figure, the target ground truth trajectory is marked with yellow circles and the estimated trajectory is marked with orange squares. Figure \ref{fig:fig4}(g) shows more clearly the agent positions at time-steps $t=[1, 3, 6, 9, 12]$, along with their sensing area which is approximated by regular dodecahedra. As can be observed, the agents position themselves in such a way so that the target is included inside their sensing area, but without causing interference to each other. The jamming interference is minimized, as shown in Fig. \ref{fig:fig4}(i), with only two jamming interference events occurring by agent $2$ towards agent $1$ during time-steps $10$ and $11$. Finally, Fig. \ref{fig:fig4}(j) shows the tracking error, which is defined here as the Euclidean distance between the estimated and ground truth target positions at each time step. The agents start with an initial target distribution which they use to predict the target state over the planning horizon. Because during the first few time steps the target is not detected, the agents are not receiving any measurements from the target, which leads to increased uncertainty over the target state and as a result the tracking error initially increases (as shown). However, as the objective of the proposed tracking-and-jamming controller is to maximize the tracking-and-jamming performance, the tracking error subsequently decreases once the agents intercept the target at $t=4$, as shown in Figs. \ref{fig:fig4}(h)-\ref{fig:fig4}(j).
\begin{figure}
	\centering
	\includegraphics[width=\columnwidth]{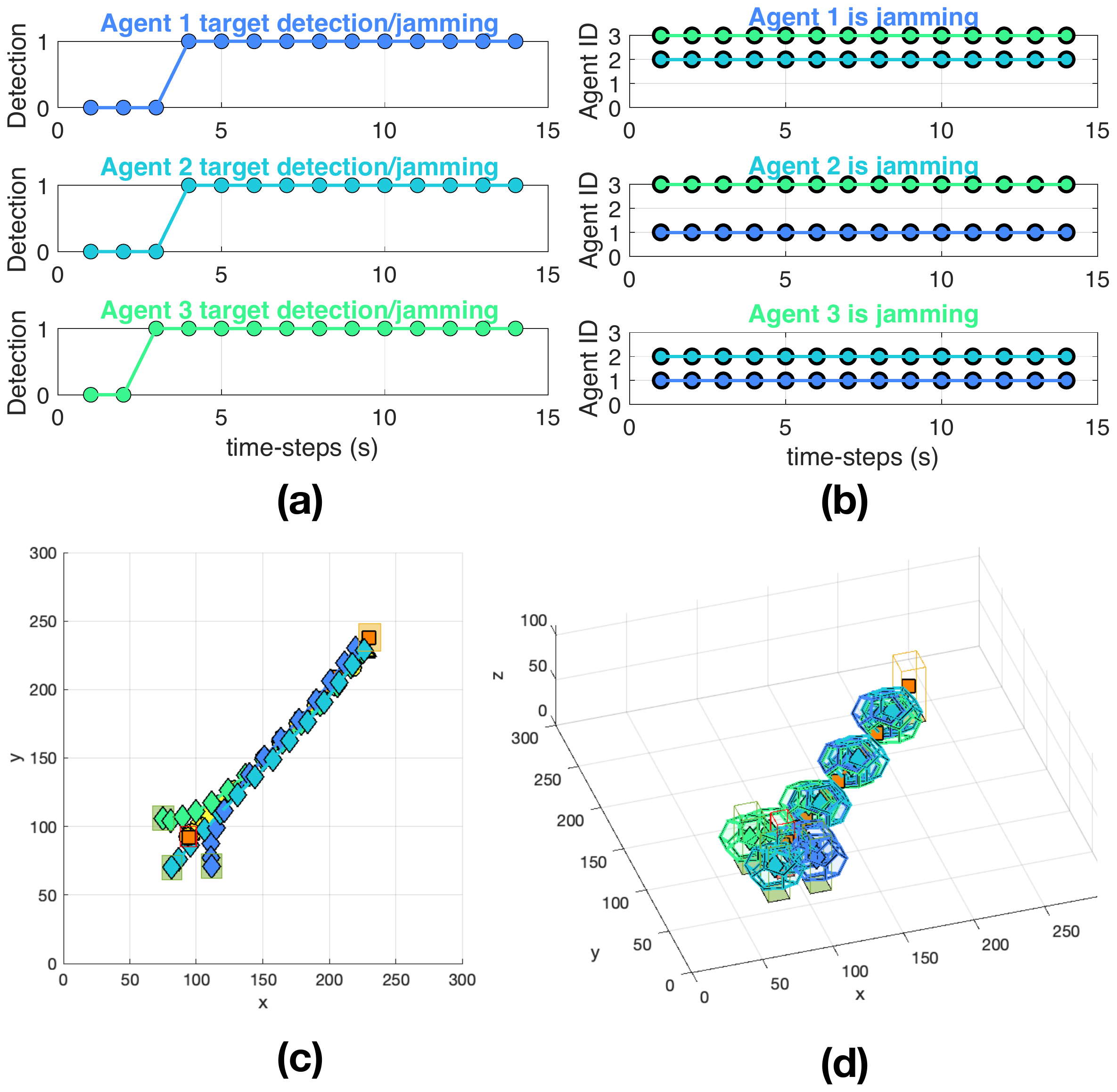}
	\caption{Concurrent tracking-and-jamming scenario with $3$ agents and $1$ target with disabled jamming interference constraints. While the target detection-and-jamming objective is maximized as shown in (a), the agents come critically close to each other, causing destructive jamming interference as shown in (b)-(d).}
	\label{fig:fig5}
	\vspace{0mm}
\end{figure}

\begin{figure}
	\centering
	\includegraphics[width=\columnwidth]{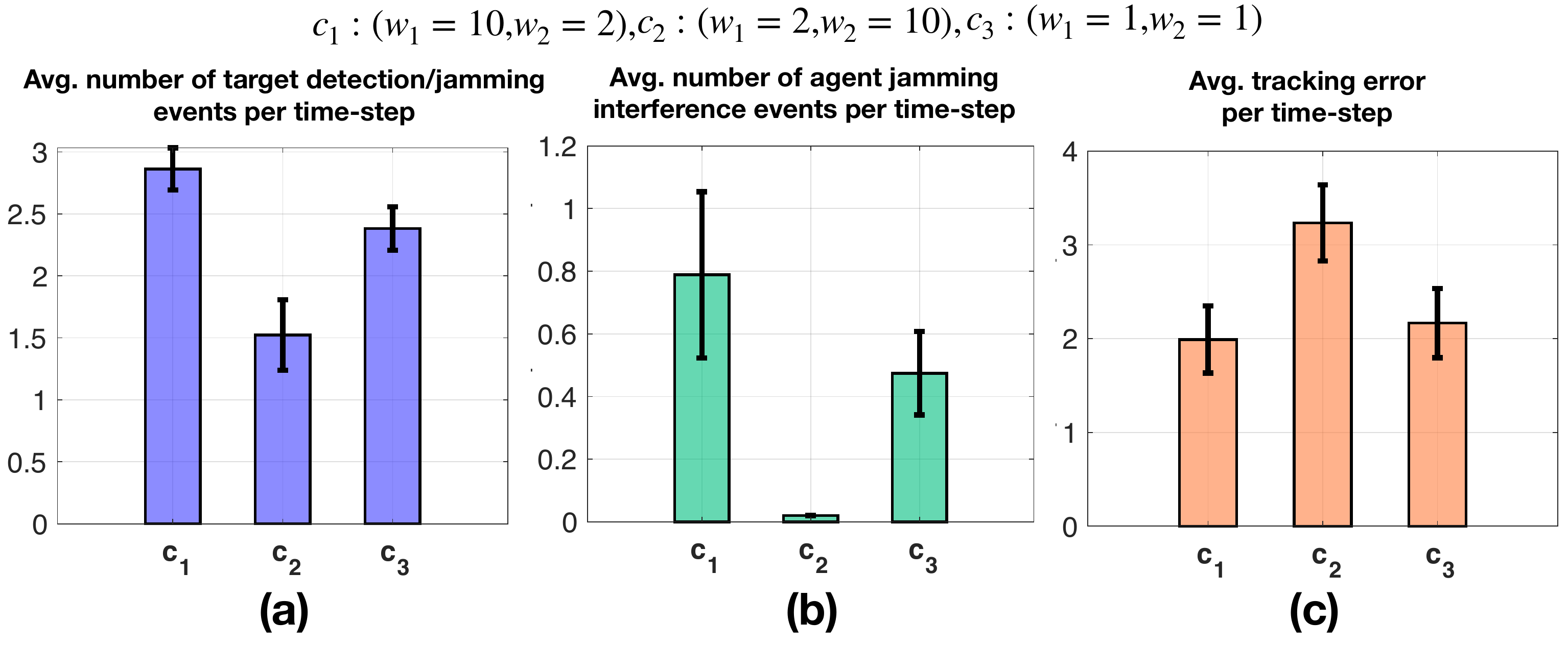}
	\caption{Effect of tuning weights $w_1$ (target detection-and-jamming objective) and $w_2$ (agent jamming interference objective) on (a) target detection-and-jamming events, (b) jamming interference events, and (c) tracking error.}
	\label{fig:fig6}
	\vspace{0mm}
\end{figure}

\begin{figure*}
	\centering
	\includegraphics[width=\textwidth]{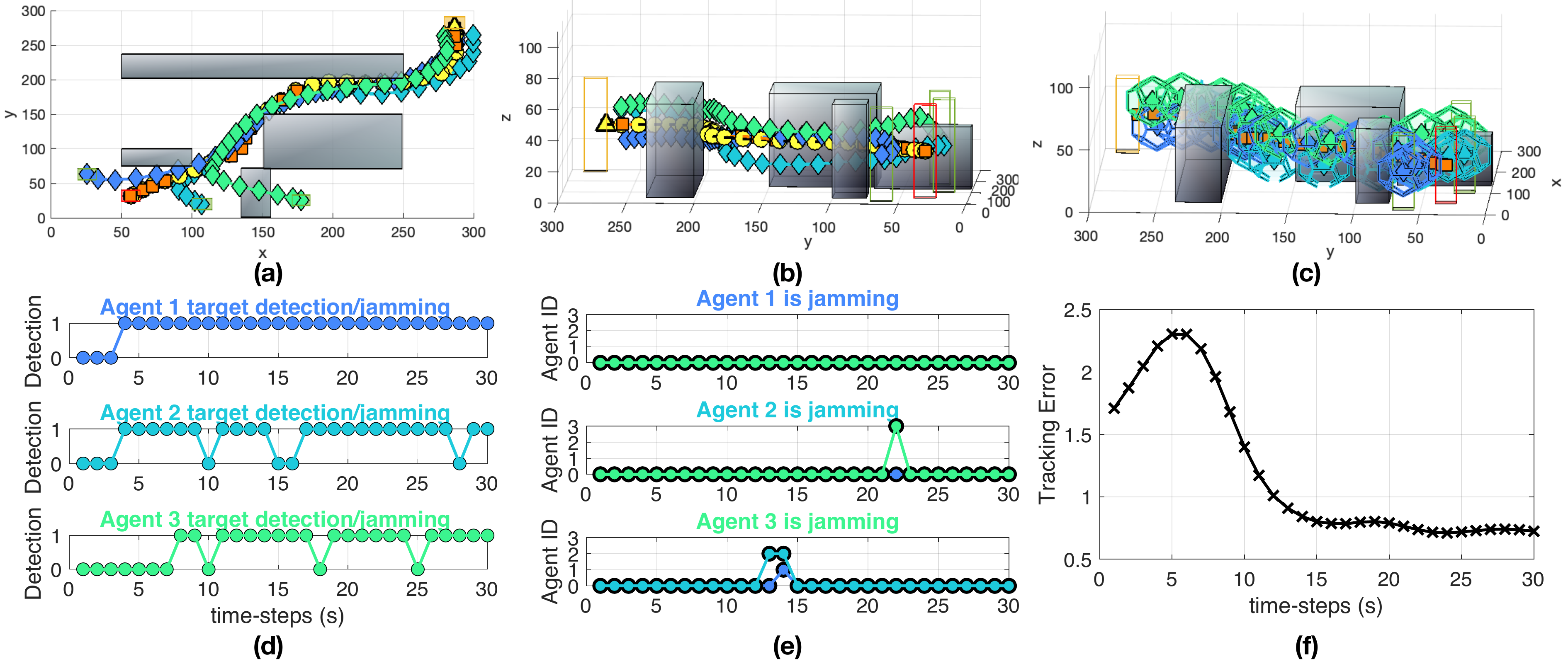}
	\caption{A tracking-and-jamming scenario with $3$ agents and $1$ target in a 3D environment with obstacles. (a)-(c) Agent and target trajectories for the duration of this experiment, (d) target detection-and-jamming events per agent, (e) jamming interference events, and (f) tracking error.}	
	\label{fig:fig7}
\end{figure*}

The next experiment illustrates the behavior of the proposed system with the jamming interference constraints disabled. In the previous experiment, the proposed tracking-and-jamming controller was executed with a weight configuration of $(w_1=1, w_2=1)$ i.e., the objective function in Eq. \eqref{eq:mi} considers both the target tracking-and-jamming events and the agent jamming interference events through weights $w_1$ and $w_2$, respectively. In this experiment, the jamming interference constraints was disabled by setting $w_2=0$ in Eq. \eqref{eq:mi} and the previous scenario was repeated ($w_3$ remained unchanged). The results of these simulations are depicted in Fig. \ref{fig:fig5}. As it can be observed, from Fig. \ref{fig:fig5}(a) the target detection-and-jamming events appear to be slightly improved compared to the previous scenario, however the jamming interference constraints are violated at every time step as shown in Fig. \ref{fig:fig5}(b). In fact, Figs. \ref{fig:fig5}(c)-\ref{fig:fig5}(d) show that the agents coalesce in their effort to maximize the target detection-and-jamming events, which causes destructive jamming interference amongst the team. Importantly, when jamming interference is accounted for (as described in the previous experiment), the proposed controller manages to achieve similar tracking-and-jamming performance as compared to the case when jamming interference is not accounted for (this experiment), while maintaining minimum jamming interference and thus keeping the system safe and operational at all times. 

In order to investigate in more detail the effect of the weighting configuration (i.e., $w_1$ and $w_2$) on the behavior of the system, a Monte Carlo (MC) simulation was performed, in which the position of the target inside the surveillance area along with the goal region were randomly initialized (the target initial location and the goal region were set to be at least $100$m apart), and then the positions of $3$ agents were uniformly sampled from a sphere with radius $25$m centered at the target location. Fifty $(50)$ trials were conducted with the system parameters set as discussed previously. In this experiment we compute a) the average number of target detection-and-jamming events per time step, b) the average number of agent jamming interference events per time step, and finally c) the average tracking error per time step, for $3$ different weight configurations $c_1=(w_1=10,w_2=2)$, $c_2=(w_1=2,w_2=10)$, and $c_3=(w_1=1,w_2=1)$. The results are shown in Fig. \ref{fig:fig6}. 

As can be observed, when the emphasis is on the target detection-and-jamming objective (i.e., $c_1$), the average number of target detection-and-jamming events is maximized as shown in Fig. \ref{fig:fig6}(a); however, as Fig. \ref{fig:fig6}(b) shows, the jamming interference events are also increased, since avoiding the jamming interference is not the main goal in this weight configuration. The opposite effect is obtained when the emphasis is on the jamming interference objective (i.e., $c_2$). In this scenario, jamming interference is minimized (Fig. \ref{fig:fig6}(b)); however, as in this configuration tracking the target is not the main goal, detection-and-jamming events are decreased. In contrast, configuration $c_3$ results in a more balanced behavior. Finally, as expected, the tracking error, shown in Fig. \ref{fig:fig6}(c) for the $3$ configurations,  closely follows the inverse trend of the target detection-and-jamming events shown in Fig. \ref{fig:fig6}(a), i.e., the more times the target is detected, the lower the tracking error. 

Next, Fig. \ref{fig:fig7} illustrates a simulated cooperative tracking-and-jamming scenario with $3$ agents and $1$ target, taking place in a bounded 3D environment with obstacles. In this scenario, agent $1$ (blue), $2$ (turquoise), and $3$ (green) are initialized at locations $[25,63,30]$, $[107,20,30]$, and $[117,25,30]$,  respectively. The target is distributed according to $\mathcal{N}(\mu_0,P_0)$, where $\mu = [55,30,30,3,3,1]$ (shown in yellow) and $P_0=\text{diag}([2, 2, 2, 0.8, 0.8, 0.8])$, and the rest of the system parameters are set as discussed in Sec. \ref{ssec:sim_setup}. The goal region, marked with a clear yellow cuboid, has centroid $[286, 282, 30]$. As can be observed from Figs. \ref{fig:fig7}(a)-\ref{fig:fig7}(b), the agents successfully track-and-jam the target, while avoiding the obstacles in their path. More specifically, as it is shown, the agents maintain a safe distance between them to avoid jamming interference, while at the same time keeping the target inside their sensing area (also shown by the target detection-and-jamming events in Fig. \ref{fig:fig7}(d)). The latter allows the agents to detect the target and acquire target measurements and thus improve the target tracking accuracy. Figure \ref{fig:fig7}(c) shows the agent positions and their sensing area over the duration of this experiment. As can be seen, the agents position themselves as close as possible to their peers, but without causing jamming interference, as depicted more clearly in Fig. \ref{fig:fig7}(e) by the sporadic jamming interference events. Finally, Fig. \ref{fig:fig7}(f) shows the tracking error for the duration of the experiment. Initially, during the time that the agents try to intercept the target, no target measurements are received, which results in increased uncertainty regarding the target state and thus increased tracking error. Once the agents detect the target (as shown in Fig. \ref{fig:fig7}(d)) and start receiving measurements, the tracking error decreases and the system's performance stabilizes, as illustrated in Fig. \ref{fig:fig7}(f).

\begin{figure}
	\centering
	\includegraphics[width=\columnwidth]{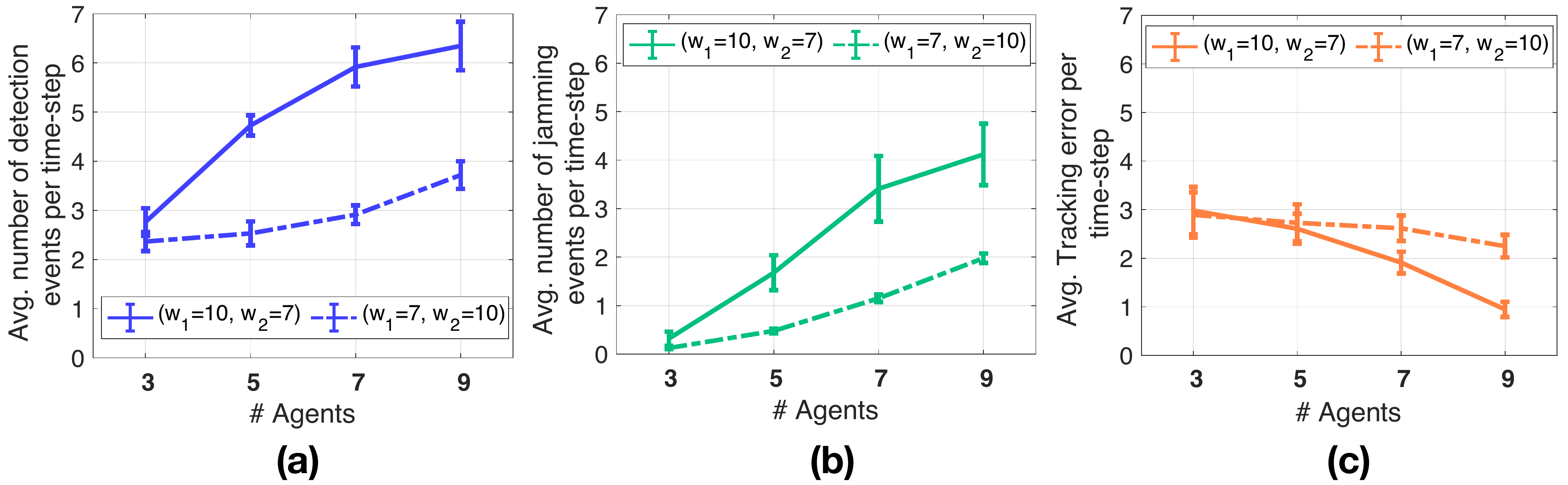}
	\caption{System performance for different number of agents and weight (i.e., $w_1$ and $w_2$)  configurations: (a) target detection-and-jamming events, (b) jamming interference events, and (c) tracking error.}
	\label{fig:fig8}
	\vspace{0mm}
\end{figure}

The next experiment aims to demonstrate how the proposed system scales with the number of agents. For this experiment, Monte Carlo simulations were conducted for $3$, $5$, $7$, and $9$ agents, randomly initialized inside the surveillance region, along with the target. The agents are spawned at locations such that their distance with the target is at maximum $30$m. Fifty $(50)$ trials were conducted for each configuration of agents and for two different weight settings i.e., $(w_1=10, w_2=7)$ and $(w_1=7, w_2=10)$, logging the target detection-and-jamming events, the agent interference events, and the tracking error, as shown in Figs. \ref{fig:fig8}(a), \ref{fig:fig8}(b), and \ref{fig:fig8}(c), respectively. When the emphasis is on the target detection-and-jamming objective i.e., $(w_1=10, w_2=7)$, it can be observed that as the number of agents increases the average number of detection-and-jamming events also increases and subsequently the tracking error decreases (Fig. \ref{fig:fig8}(c)). 

However, as shown in Fig. \ref{fig:fig8}(b), as the number of agents increases, the jamming interference events also increase. This is due to the fact that, when the number of agents increases, the number of interference-free configurations that the agents can take are reduced, i.e., there is no longer free space around the target to be occupied by the agents which results both in target detection and zero jamming interference. This issue, which is left for future work, can be alleviated in a degree by considering directional antennas and heterogeneous agents which exhibit different characteristics regarding their sensing range, sensing shape, etc. When the emphasis is on jamming interference, i.e., $(w_1=7, w_2=10)$, it can be observed from Fig. \ref{fig:fig8}(b) that the jamming interference constraints are much better addressed, with an average of just $2$ jamming interference events per time step for the case of $9$ agents. However, this occurs at the expense of target tracking-and-jamming accuracy, as shown in Figs. \ref{fig:fig8}(a)-\ref{fig:fig8}(c). Clearly, the desired system behavior (e.g., sensitivity to jamming interference, importance of tracking accuracy, etc.) can be customized and fine-tuned to match the application scenario requirements by properly selecting the weighting scheme as shown by the results of this experiment. 

%\color{blue}
\subsection{Computational Complexity}
Finally, the main factor that drives the computational complexity of the proposed approach is discussed. In general, the branch-and-bound/branch-and-cut \cite{Mitchell2002,Eckstein1996} is the main optimization method which is used to solve mixed integer quadratic programs (MIQP), and its performance is usually determined by the number of integral variables that are being utilized in the MIQP. In branch-and-bound, a search tree is constructed, which enumerates in a systematic way candidate solutions for the MIQP problem. In this tree, each node includes the original MIQP problem constraints plus additional constraints on the bounds of the integer variables (i.e., in order to enforce those variables to take integer values). The algorithm explores the nodes of this tree by optimally solving a linear programming relaxation problem (where the node's integrality constraints have been dropped). In the scenario where the optimal solution to the linear program consists of an integer constrained variable with a fractional value (i.e., $x=f$), the algorithm generates a new branch for this variable consisting of two sub-problems (i.e., nodes) where new integrality constraints are imposed (i.e., $x\le \lfloor{f}\rfloor$ and $x \ge \lceil{f}\rceil$). As the exploration proceeds, the algorithm computes and maintains lower and upper bounds on the objective function, along with the incumbent integer feasible solution, which allows it to identify and efficiently prune branches of the tree which lead to sub-optimal solutions. 

The number of integral variables (i.e., only binary variables in this work) involved in the proposed distributed controller (P3), is equal to $\mathcal{B} = b_{\tau,l} + \tilde{b}_{\tau} + \kappa_{\tau,i,l} + \tilde{\kappa}_{\tau,i} + z_{\tau,\psi,\ell}$ where $\tau = [0,..,T-1]$, $l = [1,..,12]$, $i = [1,..,|\mathcal{S}|-1]$, $\psi = [1,..,|\Psi|]$ and $\ell = [1,..,L]$. Consequently, it can be concluded that the computational complexity of the proposed controller will be affected by the length of planning horizon $T$, the number of agents $\mathcal{|S|}$, and the number of obstacles to be avoided $|\Psi|$. That said, as $T$, $\mathcal{|S|}$, and $|\Psi|$ increase, $\kappa_{\tau,i,l}, \tilde{\kappa}_{\tau,i},~\text{and}~ z_{\tau,\psi,\ell}$ are the decision variables which grow the fastest and ultimately determine the computational complexity of the controller. 
\begin{table}\normalsize
\label{tbl:tbl1}
\begin{center}
  \begin{tabular}{|c|c|c|c|}
   \hline
    \multicolumn{4}{|c|}{Avg. Execution Time (sec)} \\
    \hline
    \# Agents & Horizon Length & Centralized & Distributed\\
    \hline\hline
    3 & 3  & 0.0540  & 0.0382 \\ \hline
    5 & 3  &  0.0989 & 0.0419 \\ \hline
    7 & 3  &  1.2092 &  0.0471 \\ \hline
    3 & 5  &  0.0869  &  0.0710 \\ \hline
    5 & 5  &  1.8932  &  0.0752\\ \hline
    7 & 5  &  8.3932  &   0.0792\\ \hline
  \end{tabular}
\end{center} 
\caption{Centralized and Distributed Average Execution Time for Different (Number of agents, Horizon
Length) Combinations}
\vspace{-8mm}
\end{table}

To better understand the computational complexity of the proposed controller, a Monte-Carlo simulation was conducted, in an obstacle-free environment, where the number of agents and the length of planning horizon were varied. More specifically, for this experiment, the centralized and distributed formulation of the proposed tracking-and-jamming controller were run for $3$, $5$, and $7$ agents, with planning horizon lengths of $3$ and $5$ time-steps. For each (Number of Agents, Horizon Length) combination $20$ trials were conducted, with the agents and the target being randomly initialized inside the surveillance area. The rest of the parameters were set according to the simulation setup discussed in Sec. \ref{ssec:sim_setup}. Table~1 summarizes the results of this experiment in terms of the execution time (i.e., the time required by the solver to find the optimal solution). In particular, Table~1 shows the average time (taken over the $20$ trials) for each combination of the parameters for the centralized and distributed controllers. As can be observed from Table~1, although both the centralized and the distributed controllers seem to perform reasonably well for some values of the parameters, in terms of execution time, the distributed controller scales much better with respect to the number of agents and the length of the planning horizon. More specifically, the results show that the increasing number of agents greatly affect the computational complexity of the centralized controller as opposed to the proposed distributed approach. Observe that the problem in (P3) is a mixed integer quadratic program which contains only linear constraints with $0-1$ integral variables. These properties allow optimal solutions to be obtained on the order of milliseconds, especially with the use of the proposed distributed controller as illustrated in Table~1. It is also worth noting that additional computational savings can be obtained through various heuristics \cite{Naik2021,Hendel2021,Sonnerat2021} and approximations \cite{Klotz2013,Bertsimas2021} which can provide adequate near-optimal MIP solutions. %For instance, the optimization can be configured to stop when a feasible solution is found within the acceptable optimality level.

%However, the real-world implementation of the proposed approach and the experimentation with various heuristics and approximations that may be required for real-time operation is left for future work. In particular, our main goal for the future is to investigate the real-world limitations of the proposed approach in terms of general performance, computational complexity, and energy efficiency. To do that, we plan as a future step, to utilize our existing proof-of-concept single agent counter-drone system \cite{Souli2020} as a basis for implementing the proposed distributed multi-agent jamming controller.

\subsection{Communication Overhead}
We close the evaluation section of the proposed approach with a brief discussion on the communication requirements (i.e., amount of information exchange) of the proposed approach. In the proposed distributed estimation and control framework the agents exchange information in order to a) accurately estimate the target's trajectory over time, and b) coordinate their control decisions thus generating cooperative plans. The information exchange which takes place during the target state estimation phase is described in detail in Sec. \ref{ssec:state_estimation}. In essence, during this phase, at each time step $t$ the agents exchange their posterior belief on the target's state in a sequential fashion in an effort to compute the fused posterior distribution of the target's state. This process requires that the parameters of the target state distribution i.e., $(\mu,P)$ are transmitted in total $|S|$ times (where $|S|$ is the number of agents) amongst the agents. Since $\mu$ is a 6-by-1 column vector, and $P$ is a 6-by-6 covariance matrix, the total amount of numerical values transmitted for this purpose in each time step is equal to $42|S|$. Finally, the agents exchange their plans via a sequential processing procedure as discussed in Sec. \ref{ssec:d_control}. The total amount of information exchanged in this phase over a planning horizon of length $T$ time-steps is equal to $T|S|(|S|-1)s$, where $s$ is a 6-by-1 column vector representing the agent's state, resulting in the transmission of $6T|S|(|S|-1)$ numerical values.
\color{black}

%Moreover, we should mention here, that the proposed controller shown in (P3) can easily be converted into into a mixed-integer linear program (MILP) by setting $w_3=0$ in Eqn. \eqref{eq:mi}, without any adverse effects on the functionality of the approach. This has the potential to enable the implementation and integration of proposed approach in real-world platforms  \cite{Toupet2006,Schouwenaars2005,Culligan2007,Ragi2017,Zuo2020}.

\section{Conclusion and Future Directions} \label{sec:Conclusion}
This work investigates the problem of concurrently tracking-and-jamming a malicious target in 3D environments with a team of autonomous distributed aerial agents. A distributed model predictive control approach is proposed which allows the team of aerial agents to continuously track-and-jam the malicious drone, and at the same time minimize the jamming interference amongst them. The problem is studied in challenging scenarios with uncertain target dynamics, noisy measurements, and in the presence of obstacles that need to be avoided. The proposed approach is implemented as a mixed integer quadratic mathematical program (MIQP), which can be solved using off-the-shelf solvers. The effectiveness of the proposed approach is demonstrated through extensive simulation experiments.

Future work will utilize our existing proof-of-concept single agent counter-drone system \cite{Souli2020} as a basis for implementing the proposed distributed multi-agent jamming controller in a real-world setting, and solving possible issues which might arise from a practical implementation. For instance, the use of an omnidirectional jamming antenna might in practice interfere with the navigation circuitry of the jamming agent. A possible solution to this problem is to revert to a backup positioning system while the agent is jamming the target. Alternatively, a directional jamming antenna can be utilized. Due to battery constraints the amount of time the agents can be in the air and at the same time jam the target might be insufficient for the success of the mission; this is another issue that we plan to investigate through the practical implementation of the proposed approach. 
 
The extension of the proposed approach to multiple malicious targets will be investigated in a future step. Specifically, we will investigate the possibility of designing a pre-planning step which incorporates an instance of the assignment problem \cite{Pentico2007,Deb1992} to the proposed approach, in order to dynamically assign in an optimal way the pursuer agents to the multiple targets over a rolling time horizon. Other future directions include the investigation of the optimal number of pursuer agents that must be deployed in order to track-and-jam multiple targets within a specified bounded surveillance area and the optimization of the total time required for multiple targets to be jammed and eventually be forced to land, by multiple pursuer agents.
\color{black}

\section*{Acknowledgments}
This work has been supported by the European Union's H2020 research and innovation programme under grant agreement No 739551 (KIOS CoE - TEAMING) and from the Republic of Cyprus through the Deputy Ministry of Research, Innovation and Digital Policy.

\bibliographystyle{IEEEtran}
\bibliography{IEEEabrv,main} 

\vspace{-0.40in}

\begin{IEEEbiography}[{\includegraphics[width=1in,height=1.25in,clip,keepaspectratio]{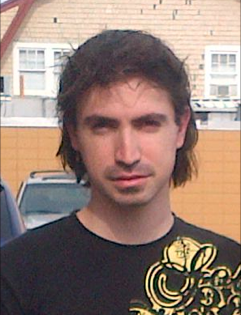}}]%
{Savvas Papaioannou} received the B.S. degree in electronic and computer engineering in 2011 from the Technical University of Crete, Greece, the M.S. degree in electrical engineering in 2013 from Yale University, USA, and the Ph.D. degree in computer science in 2017 from the University of Oxford, U.K. He is currently a Research Associate with the KIOS Research and Innovation Center of Excellence at the University of Cyprus. His research interests include multi-agent and autonomous systems, state estimation and control, multi target tracking, probabilistic inference, Bayesian reasoning, and intelligent unmanned-aircraft vehicle (UAV) systems and applications. 
\end{IEEEbiography}

\vspace{-5mm}

\begin{IEEEbiography}[{\includegraphics[width=1in,height=1.25in,clip,keepaspectratio]{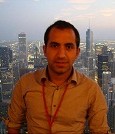}}]
{Panayiotis Kolios} is a Research Assistant Professor at the KIOS Research and Innovation Center of Excellence at the University of Cyprus. He received his B.Eng and Ph.D degrees in Telecommunications Engineering from King's College London in 2008 and 2011, respectively. His interests focus on both basic and applied research on networked intelligent systems. Some examples of such systems include intelligent transportation systems, autonomous unmanned aerial systems, and the plethora of cyber-physical systems that arise within the Internet of Things. Particular emphasis is given to emergency response aspects in which faults and attacks could cause disruptions that need to be effectively handled. He is an active member of IEEE, contributing to a number of technical and professional activities within the Association.
\end{IEEEbiography}

\vspace{-5mm}
\balance
\begin{IEEEbiography}[{\includegraphics[width=1in,height=1.25in,clip,keepaspectratio]{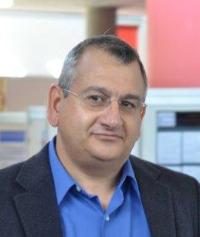}}]
{Georgios Ellinas} holds a B.S., M.Sc., M.Phil., and a Ph.D. in Electrical Engineering from Columbia University. He is a Professor and the past Chair (2014-2020) of the Department of Electrical and Computer Engineering and a founding member of the KIOS Research and Innovation Center of Excellence at the University of Cyprus. Previously, he served as Associate Professor of Electrical Engineering at City College of New York, Senior Network Architect at Tellium Inc., and Research Scientist/Senior Research Scientist at Bell Communications Research (Bellcore). Prof. Ellinas is a Fellow of the IET (2019), a Senior Member of IEEE, OSA, and ACM, and a Member of the Marie Curie Fellows Association (MCFA). His research interests are in optical networks, transportation networks, IoT, and unmanned aerial systems.
\end{IEEEbiography}

\end{document}